\documentclass[traditabstract]{aa} 
\usepackage{graphicx}
\usepackage{txfonts}
\usepackage{natbib}
\usepackage{color}
\usepackage{lscape}
\usepackage{latexsym}
\usepackage{afterpage}
\usepackage{subfigure}
\usepackage{multirow}
\usepackage{changepage}		
\bibpunct{(}{)}{;}{a}{}{,} 

\usepackage[usenames,dvipsnames,svgnames,table]{xcolor}
\usepackage[breaklinks, colorlinks, citecolor=CornflowerBlue]{hyperref}
\usepackage{url}

\begin{document}
\definecolor{Red}{rgb}{1,0,0}
\authorrunning{Remco F.J. van der Burg et al.}
   \title{Prospects for high-$z$ cluster detections with \textit{Planck}, based on a follow-up of 28 candidates using MegaCam@CFHT}
   \titlerunning{A follow-up study of 28 high-$z$ \textit{Planck} cluster candidates using MegaCam@CFHT}	
    \author{R.~F.~J.~van der Burg\inst{1}, H.~Aussel\inst{1}, G.~W.~Pratt\inst{1}, M.~Arnaud\inst{1}, J.-B.~Melin\inst{2}, N.~Aghanim\inst{3}, R.~Barrena\inst{4,5}, H.~Dahle\inst{6}, M.~Douspis\inst{3}, A.~Ferragamo\inst{4,5}, S.~Fromenteau\inst{7}, R.~Herbonnet\inst{8}, G.~Hurier\inst{3,9}, E.~Pointecouteau\inst{10,11}, J.~A.~Rubi\~no-Mart\'in\inst{4,5}, A.~Streblyanska\inst{4,5}}
	   \institute{Laboratoire AIM, IRFU/Service d'Astrophysique - CEA/DSM - CNRS - Universit\'e Paris Diderot, B\^at. 709, CEA-Saclay, 91191 Gif-sur-Yvette Cedex, France\\
                    \email{remco.van-der-burg@cea.fr}  
                    \and DSM / IRFU / SPP, CEA-Saclay, F-91191 Gif-sur-Yvette Cedex, France
                    \and Institut d'Astrophysique Spatiale, CNRS (UMR8617) Universit\'e Paris-Sud 11, B\^atiment 121, Orsay, France
                    \and Instituto de Astrof\'isica de Canarias, C/V\'ia L\'actea s/n, La Laguna, Tenerife, Spain
                    \and Dpto. Astrof\'isica, Universidad de La Laguna (ULL), E-38206 La Laguna, Tenerife, Spain   
                    \and Institute of Theoretical Astrophysics, University of Oslo, P.O. Box 1029, Blindern, N-0315 Oslo, Norway
				    \and Departments of Physics, Carnegie Mellon University, 5000 Forbes Ave., Pittsburgh, PA 15217, USA	                    
                    \and Leiden Observatory, Leiden University, PO Box 9513, 2300 RA Leiden, the Netherlands
                    \and Laboratoire de Physique Subatomique et de Cosmologie, Universit\'e Joseph Fourier Grenoble I, CNRS/IN2P3, Institut National Polytechnique de Grenoble, 53 rue des Martyrs, 38026 Grenoble cedex, France
                    \and Universit\'e de Toulouse, UPS-OMP, IRAP, F-31028 Toulouse cedex 4, France
                    \and CNRS, IRAP, 9 Av. colonel Roche, BP 44346, F-31028 Toulouse cedex 4, France
                       }             
             
   \date{Submitted September 3rd 2015; accepted December 1st 2015}

  \abstract{The \textit{Planck} catalogue of SZ sources limits itself to a significance threshold of 4.5 to ensure a low contamination rate by false cluster candidates. This means that only the most massive clusters at redshift $z>0.5$, and in particular $z>0.7$, are expected to enter into the catalogue, with a large number of systems in that redshift regime being expected around and just below that threshold. In this paper, we follow-up a sample of SZ sources from the \textit{Planck} SZ catalogues from 2013 and 2015. In the latter maps, we  consider detections around and at lower significance than the threshold adopted by the \textit{Planck} Collaboration. To keep the contamination rate low, our 28 candidates are chosen to have significant WISE detections, in combination with non-detections in SDSS/DSS, which effectively selects galaxy cluster candidates at redshifts $z\gtrsim 0.5$. By taking $r$- and $z$-band imaging with MegaCam@CFHT, we bridge the 4000$\mathrm{\AA}$ rest-frame break over a significant redshift range, thus allowing accurate redshift estimates of red-sequence cluster galaxies up to $z\sim 0.8$. After discussing the possibility that an overdensity of galaxies coincides -by chance- with a \textit{Planck} SZ detection, we confirm that 16 of the candidates have likely optical counterparts to their SZ signals, 13 (6) of which have an estimated redshift $z>0.5$ ($z>0.7$). The richnesses of these systems are generally lower than expected given the halo masses estimated from the \textit{Planck} maps. However, when we follow a simplistic model to correct for Eddington bias in the SZ halo mass proxy, the richnesses are consistent with a reference mass-richness relation established for clusters detected at higher significance. This illustrates the benefit of an optical follow-up, not only to obtain redshift estimates, but also to provide an independent mass proxy that is not based on the same data the clusters are detected with, and thus not subject to Eddington bias.}  
   \maketitle
%

\hyphenation{in-tra-clus-ter}
\hyphenation{rank-or-der}

\section{Introduction}

Our fiducial Lambda-Cold-Dark-Matter ($\Lambda$CDM) cosmological paradigm provides a model in which structures form hierarchically. The most massive gravitationally collapsed systems in this picture, called galaxy clusters, provide insights into the physics at play in the extremely hot and highly ionised Intra-Cluster Medium (ICM), constitute unique laboratories to study the evolution of galaxies as a function of their environment, and are of particular interest for understanding structure formation through the statistical study of the cluster population. Furthermore, the abundance of galaxy clusters, as a function of mass and redshift, is sensitive to the underlying cosmology \citep[e.g.][and references therein]{allen11}, and thus provides a means to verify the cosmological model \citep[e.g.][]{vikhlinin09,rozo10,mantz10a,sehgal11,benson13,planck15clustercosmology}.

An observationally selected mass-limited sample of galaxy clusters would be ideal, as it would allow for a straightforward comparison with theoretical predictions when probing cluster formation physics or cosmology.
In practice, such a selection cannot be made, since halo mass is not a direct observable. Historically, baryonic tracers such as galaxies and X-ray emitting gas have been used as mass-proxies to construct samples of galaxy clusters \citep[e.g.][]{abell89,ebeling98,gladdersyee00,bohringer00,bohringer04,gilbank11,gettings12}. However, due to differences in assembly histories, and the complexity of the physics involved, such detection methods potentially bias a cluster sample towards a specific dynamical state, or are particularly subject to projection effects along the line-of-sight. Cluster samples constructed in such ways still require significant corrections to link the observables to the halo mass, before comparing the data to theoretical predictions. This also complicates a cosmological analysis based on the abundance of clusters, since that requires a precise knowledge of the selection function and catalogue completeness. 

A detection method proposed several decades ago,  based on the Sunyaev-Zeldovich \citep[][SZ]{sunyaev80} effect, is expected to yield cluster samples that are  closer to being mass-selected \citep[e.g.][]{dasilva04,hoekstra12}. Over the last few years, SZ-based cluster samples have been provided over areas of 100s to 1000s of square degrees by the South Pole Telescope \citep[SPT,][]{reichardt13,bleem15}, and the Atacama Cosmology Telescope \citep[ACT,][]{hasselfield13}. The first all-sky SZ-detected cluster catalogue is provided by \textit{Planck}, which released three catalogues during the course of its lifetime. The first contained 189 clusters and was based on about a third of the final data set \citep{PlanckESZ}. A more extensive catalogue, containing SZ detections down to a significance of S/N=4.5, was released after about half of the survey lifetime in 2013 \citep[][hereafter PSZ1]{psz1}. The final \textit{Planck} maps and SZ catalogues, based on the full mission, were published recently \citep[][hereafter PSZ2]{psz2}. 

In practice, any catalogue down to sufficiently low S/N includes false detections. Although the average purity can be estimated statistically \citep[]{psz1,psz2}, a full exploitation of the catalogue requires one to identify bona-fide clusters and to measure their redshifts. By cross-matching \textit{Planck} SZ detections with existing cluster catalogues, and by identifying galaxy overdensities in existing survey data around the SZ sources, the \textit{Planck} Collaboration has confirmed clusters as part of their analysis \citep{planckpsz1v215,psz2}. As described in detail in these papers, the catalogue validation process also includes the results from several studies that are published independently \citep[e.g.][]{rozo15,liu15} and dedicated validation follow-up \citep[e.g.][]{planckxmm11, planckrtt15}. This verification work is still on-going, and is now focussed on a systematic follow-up of remaining SZ candidates, such as the recent work of \citet{planckcanary15}. Since the purity decreases (i.e. the number of false candidates increases) with decreasing S/N, systematic follow-up is less and less efficient at unravelling new clusters. Another difficulty, which is more severe at low S/N, is the ambiguity in SZ source confirmation, i.e. in assessing whether a galaxy overdensity is the real counterpart, or a chance association that is spatially coincident with the $\sim\,5\arcmin$ \textit{Planck} beam. 

Nevertheless, there is an interest to go to even lower S/N than the published Planck catalogues. For instance, going to lower S/N than a given limit may help to understand the completeness above that limit. More importantly for the present paper, the all-sky nature of the \textit{Planck} catalogue allows us to probe the rarest objects that populate the high-mass end of the mass function. This is a unique capability of \textit{Planck}, of particular importance in the high-redshift regime. The previous  All-Sky Survey, performed in X--ray by the ROSAT satellite, had a limited depth and even the MACS survey \citep{ebeling07}, based on a systematic follow-up of the ROSAT Bright Source Catalogue, detected only 12 objects at $z > 0.5$, all of which lie at $z<0.7$.  
However, the prospects to construct {\it sizeable} samples of massive clusters ($M_{500}\gtrsim5\times10^{14}\mathrm{M_{\odot}}$)\footnote{All quoted masses in this paper are defined with respect to the critical density at the cluster redshift. $R_{500}$ is thus defined to be the radius at which the mean interior density is 500 times the critical density, and $M_{500}$ is the mass contained within this radius.}  spread  over a range of high redshifts ($0.5<z<1.0$),
is still limited.
With a significance threshold of S/N=4.5, the published PSZ2 catalogue is $\sim 80\%$ complete for $M_{500}\gtrsim7.5\times10^{14}\mathrm{M_{\odot}}$ at $z\gtrsim0.5$, but its  completeness decreases to $20\%$ for $M_{500}\gtrsim5.0\times10^{14}\mathrm{M_{\odot}}$ \citep[Fig.~26 in][]{psz2}.
By lowering the significance threshold, one quickly gains in completeness and thus unveils more of these high-$z$ massive clusters, a
 gain amplified by the fact that we are at the exponential end of the halo mass function at these redshifts. 
The main challenge is that  these clusters have to be identified among an increasing number of candidates, the majority being at low redshift, together with an increasing fraction of false candidates as the detection significance (i.e., purity) decreases. A fully systematic follow-up of all these candidates would not only be very inefficient, but no longer feasible in practice. However, in this paper we explore the use of existing optical and near-infrared survey data to pre-select likely massive high-$z$ clusters among the candidates, before performing the deeper follow-up observations. 

This study focuses on a sample of 28 cluster candidates, which are either part of the PSZ1 catalogue, the PSZ2 catalogue, or are detected at lower SZ significance in the final \textit{Planck} maps. It therefore contributes to the general systematic validation of \textit{Planck} cluster candidates in the public catalogues, but also takes a first step towards the construction of a sample of massive high-$z$ samples beyond the standard detection limit. Specifically, this is a pilot study to (1) investigate how existing (optical and near-infrared) survey data allow us to study detections at lower SZ significance, while maintaining a high purity of detecting real clusters in general, and redshift $z\gtrsim 0.5$ clusters in particular, (2) illustrate the importance of a \textit{quantitative} way to characterise optical counterparts of cluster candidates in deeper follow-up data to verify if it is expected for a halo corresponding to the measured SZ signal, and related to this (3) study the effect of Eddington bias on the SZ mass proxy at low detection significance.

The present work concerns an inhomogeneously selected sample of cluster candidates that does not have a clearly defined selection function. As such it is inappropriate for cosmological studies. The eventual goal of the project is to obtain a representative sample of the most massive clusters at $z > 0.5$ (and particularly $z > 0.7$). Representativity is  key for the study of the statistical properties of clusters (e.g. their baryon fractions and profiles, or their total mass profiles), as a probe of the physics of structure formation. In this context, a sample does not have to be complete or to have a precisely quantified selection function, such as is necessary for cosmological applications, so long as it is representative of the underlying population.

\afterpage{
\clearpage
\begin{landscape}
\begin{table}
\caption{\footnotesize The sample of 28 cluster candidates studied here. Listed first are confirmed candidates, below which are listed invalidated candidates.}
\begin{center}
\begin{adjustwidth}{-0.8cm}{}
\begin{tabular}{clcccccccccclr}
\hline \hline 
&&PSZ1$_{\mathrm{ID}}$&PSZ2$_{\mathrm{ID}}$&&&Distance&&&&& SZ Mass$^{\mathrm{c}}$ &&Richness Mass$^{\mathrm{d}}$\\
&Name & (\texttt{CLASS,QN})&(\texttt{Q\_NEURAL})& SNRblind$^{\mathrm{a}}$&SNRre-ext$^{\mathrm{a}}$& [arcmin] &RA$_\mathrm{J2000}^{\mathrm{b}}$ & Dec$_\mathrm{J2000}^{\mathrm{b}}$ &$z_\mathrm{RS}$ &$z_\mathrm{ref}$&[$10^{14}\, \mathrm{M_{\odot}}$]&$\,\,\,\,\,\,\,\,$Richness$^{\mathrm{d}}$&[$10^{14}\, \mathrm{M_{\odot}}$]\\
\hline
\parbox[t]{2mm}{\multirow{16}{*}{\rotatebox[origin=c]{90}{Confirmed}}}&\texttt{PLCK G027.65-34.27}&-&-&3.68&3.68&0.00&20:49:37.9&$-$18:55:57.6&$0.58_{-0.02}^{+0.03}$&-&$5.14_{-0.91}^{+0.82}$&$\,\,\,41.6\pm8.9$&$2.18\pm0.48$\\
&\texttt{PLCK G038.64-41.15}&-&-&3.64&3.42&1.72&21:29:43.2&$-$13:28:57.0&$0.56_{-0.03}^{+0.03}$&-&$5.02_{-0.96}^{+0.86}$&$\,\,\,45.3\pm8.8$&$2.38\pm0.48$\\
&\texttt{PSZ2 G041.69+21.68}&116(1,0.99)&151(0.99)&4.33&4.09&1.72&17:47:12.2&+17:10:33.3&$0.47_{-0.03}^{+0.04}$&0.479$^{\mathrm{h}}$&$5.23_{-0.85}^{+0.77}$&$\,\,\,70.8\pm11.5$&$3.79\pm0.64$\\
&\texttt{PSZ2 G042.32+17.48}&117(1,0.96)&153(0.99)&4.93&4.42&2.43&18:04:16.5&+16:02:20.9&$0.48_{-0.03}^{+0.03}$&0.458$^{\mathrm{f}}$&$5.52_{-0.82}^{+0.75}$&$\,\,\,83.8\pm13.2$&$4.50\pm0.74$\\
&\texttt{PSZ2 G048.21-65.00}&150(1,0.88)&191(0.93)&5.15&4.96&2.43&23:09:51.0&$-$18:19:56.9&$0.41_{-0.02}^{+0.03}$&0.407$^{\mathrm{g,l}}$&$5.35_{-0.73}^{+0.67}$&$\,\,\,68.6\pm10.2$&$3.66\pm0.56$\\
&\texttt{PSZ2 G071.82-56.55}&-&304(0.86)&4.47&4.47&0.00&23:09:35.2&$-$04:09:59.9&$0.87_{-0.04}^{+0.06}$&-&$5.87_{-0.81}^{+0.75}$&$143.2\pm13.8$&$7.85\pm0.79$\\
&\texttt{PSZ2 G076.18-47.30}&-&324(0.85)&5.30&4.81&2.43&22:52:35.2&+04:32:27.0&$0.72_{-0.04}^{+0.02}$&0.666$^{\mathrm{i}}$&$5.62_{-0.89}^{+0.80}$&$142.8\pm13.8$&$7.83\pm0.79$\\
&\texttt{PLCK G079.95+46.96}&-&-&4.12&3.42&2.43&16:02:11.7&+51:03:45.1&$0.79_{-0.08}^{+0.04}$&-&$4.36_{-0.80}^{+0.72}$&$\,\,\,40.7\pm8.5$&$2.13\pm0.46$\\
&\texttt{PLCK G087.58-41.63}&-&-&3.22&2.87&1.72&23:05:43.9&+13:52:35.0&$0.98_{-0.09}^{+0.00}$&-&$3.56_{-1.60}^{+1.15}$&$\,\,\,60.3\pm16.2$&$3.21\pm0.89$\\
&\texttt{PSZ2 G106.15+25.75}&383(1,0.94)&513(0.92)&4.31&4.31&0.00&18:56:51.9&+74:55:53.4&$0.63_{-0.03}^{+0.03}$&0.588$^{\mathrm{g}}$&$4.60_{-0.74}^{+0.67}$&$\,\,\,44.4\pm9.2$&$2.34\pm0.50$\\
&\texttt{PSZ2 G119.30-64.68}&-&586(0.96)&5.62&5.31&1.72&00:45:12.5&$-$01:52:31.6&$0.57_{-0.03}^{+0.02}$&0.557$^{\mathrm{i,j}}$&$6.58_{-0.80}^{+0.74}$&$105.2\pm12.4$&$5.71\pm0.70$\\
&\texttt{PSZ2 G141.77+14.19}&508(1,0.96)&689(0.94)&5.37&5.18&1.72&04:41:05.2&+68:13:21.9&$0.77_{-0.08}^{+0.04}$&0.821$^{\mathrm{k}}$&$7.82_{-0.98}^{+0.91}$&$\,\,\,90.6\pm20.4$&$4.89\pm1.14$\\
&\texttt{PLCK G191.75-21.78}&-&-&4.54&3.40&2.43&04:54:49.1&+07:28:22.3&$0.60_{-0.03}^{+0.05}$&-&$5.78_{-1.13}^{+1.01}$&$103.4\pm12.7$&$5.60\pm0.72$\\
&\texttt{PSZ2 G198.80-57.57}&-&902(0.96)&3.66&3.65&1.72&03:02:06.5&$-$15:33:31.6&$0.55_{-0.04}^{+0.03}$&-&$5.15_{-0.93}^{+0.83}$&$\,\,\,45.8\pm8.6$&$2.41\pm0.47$\\
&\texttt{PSZ2 G208.57-44.31}&-&937(0.95)&4.26&4.13&1.72&04:02:35.4&$-$15:40:55.0&$0.85_{-0.07}^{+0.02}$&-&$5.99_{-0.92}^{+0.84}$&$\,\,\,50.8\pm10.1$&$2.68\pm0.55$\\
&\texttt{PLCK G227.99+38.11}&-&-&3.66&2.19&1.72&09:32:21.9&+05:41:02.1&$0.81_{-0.04}^{+0.06}$&-&$4.03_{-1.25}^{+1.04}$&$\,\,\,58.0\pm9.5$&$3.08\pm0.52$\\
\hline
\parbox[t]{2mm}{\multirow{12}{*}{\rotatebox[origin=c]{90}{Invalidated}}}&\texttt{PSZ1 G023.38-33.46}&58(2,0.93)&-&2.21&1.87&1.72&20:41:14.5&$-$21:55:40.3&$0.79_{-0.02}^{+0.12}$&-&$3.42_{-1.50}^{+1.12}$&$\,\,\,\,\,\,3.1\pm5.4$&$0.15\pm0.27$\\
&\texttt{PSZ1 G031.41+28.75}&84(2,0.99)&-&3.97&1.92&3.84&17:04:47.4&+11:28:12.7&$0.42_{-0.05}^{+0.02}$&-&$3.07_{-1.32}^{+1.00}$&$\,\,\,20.9\pm7.4$&$1.07\pm0.40$\\
&\texttt{PSZ2 G037.67+15.71}&102(1,0.00)&135(0.00)&6.61&6.61&0.00&18:03:13.9&+11:12:14.7&$0.60_{-0.06}^{+0.07}$&-&$6.93_{-0.84}^{+0.77}$&$\,\,\,24.1\pm11.5$&$1.24\pm0.62$\\
&\texttt{PSZ1 G038.25-58.36}&104(2,0.90)&-&1.40&0.95&1.72&22:36:07.0&$-$20:09:11.1&$0.67_{-0.02}^{+0.03}$&-&$2.19_{-0.00}^{+1.29}$&$\,\,\,37.7\pm8.2$&$1.97\pm0.44$\\
&\texttt{PSZ1 G051.42-26.16}&162(2,0.78)&-&1.87&1.54&1.72&20:57:26.0&+03:01:33.4&$0.92_{-0.11}^{+0.02}$&-&$3.36_{-1.57}^{+1.13}$&$\,\,\,30.8\pm14.2$&$1.60\pm0.77$\\
&\texttt{PLCK G053.41+61.50}&-&-&3.04&1.41&3.44&15:00:30.1&+33:18:45.9&$0.72_{-0.06}^{+0.10}$&-&$3.02_{-1.54}^{+1.10}$&$\,\,\,10.2\pm4.9$&$0.51\pm0.25$\\
&\texttt{PSZ1 G053.50+09.56}&165(2,0.00)&-&2.95&1.53&2.43&18:53:59.0&+22:30:59.3&$0.12_{-0.02}^{+0.03}$&-&$3.32_{-1.61}^{+1.16}$&$\,\,\,\,\,\,2.1\pm6.6$&$0.10\pm0.33$\\
&\texttt{PSZ2 G071.67-42.76}&239(2,0.00)&303(0.01)&8.37&7.81&3.44&22:30:45.7&+05:40:30.8&$0.77_{-0.03}^{+0.12}$&-&$7.56_{-0.75}^{+0.70}$&$\,\,\,12.0\pm5.8$&$0.60\pm0.30$\\
&\texttt{PSZ1 G081.56+31.03}&271(2,0.87)&-&3.11&3.06&1.72&17:45:53.4&+53:49:43.1&$0.76_{-0.09}^{+0.13}$&-&$4.48_{-0.92}^{+0.83}$&$\,\,\,13.1\pm5.7$&$0.66\pm0.30$\\
&\texttt{PSZ1 G092.41-37.39}&317(1,0.00)&-&2.43&2.00&1.72&23:10:15.2&+19:21:41.9&$0.22_{-0.12}^{+0.04}$&0.114$^{\mathrm{f}}$&$1.98_{-0.89}^{+0.77}$&$\,\,\,10.7\pm4.8$&$0.53\pm0.25$\\
&\texttt{PSZ2 G157.07-33.63}&549(2,0.18)&757(0.07)&5.08&4.81&1.72&02:51:34.1&+21:08:08.9&$0.87_{-0.04}^{+0.01}$&-&$6.90_{-1.01}^{+0.94}$&$\,\,\,20.6\pm10.7$&$1.05\pm0.57$\\
&\texttt{PSZ1 G240.42+77.58}&809(2,0.95)&-&2.96&2.22&3.44&12:04:14.9&+20:57:33.8&$0.57_{-0.09}^{+0.06}$&-&$3.50_{-1.19}^{+0.97}$&$\,\,\,\,\,\,2.2\pm4.0$&$0.11\pm0.20$\\
\hline
\label{tab:overview}
\end{tabular}
\end{adjustwidth}
\end{center}
\begin{list}{}{}
\item[$^{\mathrm{a}}$] Signal-to-noise ratio of the SZ signal using the MMF3 detection pipeline on the final \textit{Planck} maps, for a blind search (SNRblind), or fixing the position to the galaxy overdensity (SNRre-ext). 
\item[$^{\mathrm{b}}$] Location that maximizes the richness measurement. 
\item[$^{\mathrm{c}}$] SZ halo mass proxy ($M_{500}$) following Arnaud et al. (in prep) at the location of the galaxy overdensity.
\item[$^{\mathrm{d}}$] Richness estimator and associated mass proxy from \citet{rozo15}. 
\item[$^{\mathrm{f}}$] ENO paper \citep{planckcanary15}, also mentioned in \citet{planckpsz1v215}.
\item[$^{\mathrm{g}}$] Redshifts with the Russian-Turkish Telescope \citep{planckrtt15}.
\item[$^{\mathrm{h}}$] Overlapping with Pan-STARRS \citep{liu15}.
\item[$^{\mathrm{i}}$] SDSS BOSS spectra \citep{sdssdr12}.
\item[$^{\mathrm{j}}$] ACT cluster \citep{kirk15}.
\item[$^{\mathrm{k}}$] This spectroscopic redshift was obtained using telescope time awarded by the CCI International Time Programme at the Canary Islands Observatories (program ITP13-8, PI: Rubino-Martin).
\item[$^{\mathrm{l}}$] Spectroscopic redshift presented in Dahle et al., (in prep.).
\end{list}
\end{table}
\clearpage
\end{landscape}}

The structure of this paper is as follows. The cluster candidate sample we considered for follow-up is presented in Sect.~\ref{sec:sample}, while the follow-up data and optical catalogues are presented in Sect.~\ref{sec:data}. In Sect.~\ref{sec:analysis} we describe the red-sequence model that we use to find galaxy overdensities close to the \textit{Planck} SZ detections, and measure their redshifts and richnesses. We discuss the likelihood that these counterparts are truly associated with a given SZ detection in Sect.~\ref{sec:counterparts}, and discuss the relation between mass and richness for this sample. In particular, we discuss the effect of Eddington bias in the SZ mass proxy in Sect.~\ref{sec:eddington}, which is important to interpret our measured mass-richness relation. Sect.~\ref{sec:candidates} contains a discussion of individual candidates, for which pseudo-colour images are shown in Appendix~\ref{sec:images}. We summarise and conclude in Sect.~\ref{sec:summary}.

All magnitudes we quote are in the AB magnitudes system, and we adopt $\Lambda$CDM cosmology with $\Omega_{\mathrm{m}}=0.3$, $\Omega_{\Lambda}=0.7$ and $\mathrm{H_0=70\, km\, s^{-1}\,  Mpc^{-1}}$.

\section{\textit{Planck} sample}\label{sec:sample}
\subsection{Candidate selection}
The \textit{Planck} PSZ1 catalogue consists of extended sources detected at a significance of S/N$>$4.5 in the first release (i.e. based on about half of the final data set). Candidates which were not yet validated to be a cluster at that time were sorted into \texttt{CLASS1-3} according to their likelihood of being a real cluster (from high to low). This classification scheme was based on an SZ-quality assessment combined with information from external data from the Rosat All Sky Survey (RASS) and the Wide-field Infrared Survey Explorer \citep[WISE,][]{wise}. In semesters 2013A and 2013B we targeted a total of 16 \texttt{CLASS1} and \texttt{CLASS2} candidates in the Northern hemisphere ($\delta_{\mathrm{J2000}} > -25\degr$) with MegaCam. Note that the follow-up of these candidates is part of a larger validation program which aims at verifying \textit{all} candidates of the PSZ1 catalogue. The ones we pursue in this paper were not (yet) confirmed to be actual clusters at that time, and were picked because they were  possibly $z\gtrsim 0.5$ clusters as they did not show any obvious counterpart in the Digitized Sky Survey (DSS\footnote{The Digitized Sky Surveys were produced at the Space Telescope Science Institute under U.S. Government grant NAG W-2166. The images of these surveys are based on photographic data obtained using the Oschin Schmidt Telescope on Palomar Mountain and the UK Schmidt Telescope. The plates were processed into the present compressed digital form with the permission of these institutions.}), nor in SDSS (where available). The targets are presented in Table~\ref{tab:overview}, where PSZ1 entry numbers and associated classifications are listed.

During semester 2014B, we targeted another 12 candidates with MegaCam, this time selected from the final maps and SZ catalogue (PSZ2). We improved our preferential selection of high-$z$ cluster candidates for semester 2014B by combining information from WISE and the DSS in the following way. Massive cluster galaxies with redshifts $z\gtrsim 0.5$ are expected to be significantly detected in the WISE 3.4$\mu$m channel, while showing no significant detection in the relatively shallow optical DSS images \citep[e.g.][]{fassbender11}. We exploit this information by searching for overdensities of such galaxies within $\sim 4'$ from all \textit{Planck} SZ detections down to a significance of S/N$>$4.0 in the final maps, which did not correspond to known clusters. This way we selected, by visual inspection, 12 candidates in the Northern hemisphere ($\delta_{\mathrm{J2000}} > -25\degr$) that are likely coincident with a system of high-$z$ galaxies. An overview of the full list of targeted candidates is given in Table~\ref{tab:overview}, with entries in the PSZ1 and PSZ2 catalogues (if applicable). 

The PSZ2 catalogue, and also the updated version of the PSZ1 catalogue \citep{planckpsz1v215}, contain a classification of SZ detections based on a supervised neural network. As described in \citet{aghanim15}, the quality flag \texttt{Q\_NEURAL} (or \texttt{QN} for PSZ1) provides a condensed 1-dimensional description of the contribution of components other than the SZ effect to the SED measured by \textit{Planck}. Physical sources of contamination include the CMB, infrared emission from Galactic dust, molecular Galactic CO emission, and a radio component from Galactic free-free, synchrotron, and thermal dust emission. A high value of \texttt{Q\_NEURAL}$\gtrsim$0.4 indicates that a source has an SED dominated by the SZ effect, whereas a lower value suggest a distorted SED, and thus a likely false candidate. We list \texttt{Q\_NEURAL} parameters from \citet{aghanim15} in Table~\ref{tab:overview}, and discuss our candidates in this context in Sect.~\ref{sec:yesorno}.

\begin{figure*}
\centering
\begin{minipage}{.5\textwidth}
  \centering
  \includegraphics[width=.95\linewidth]{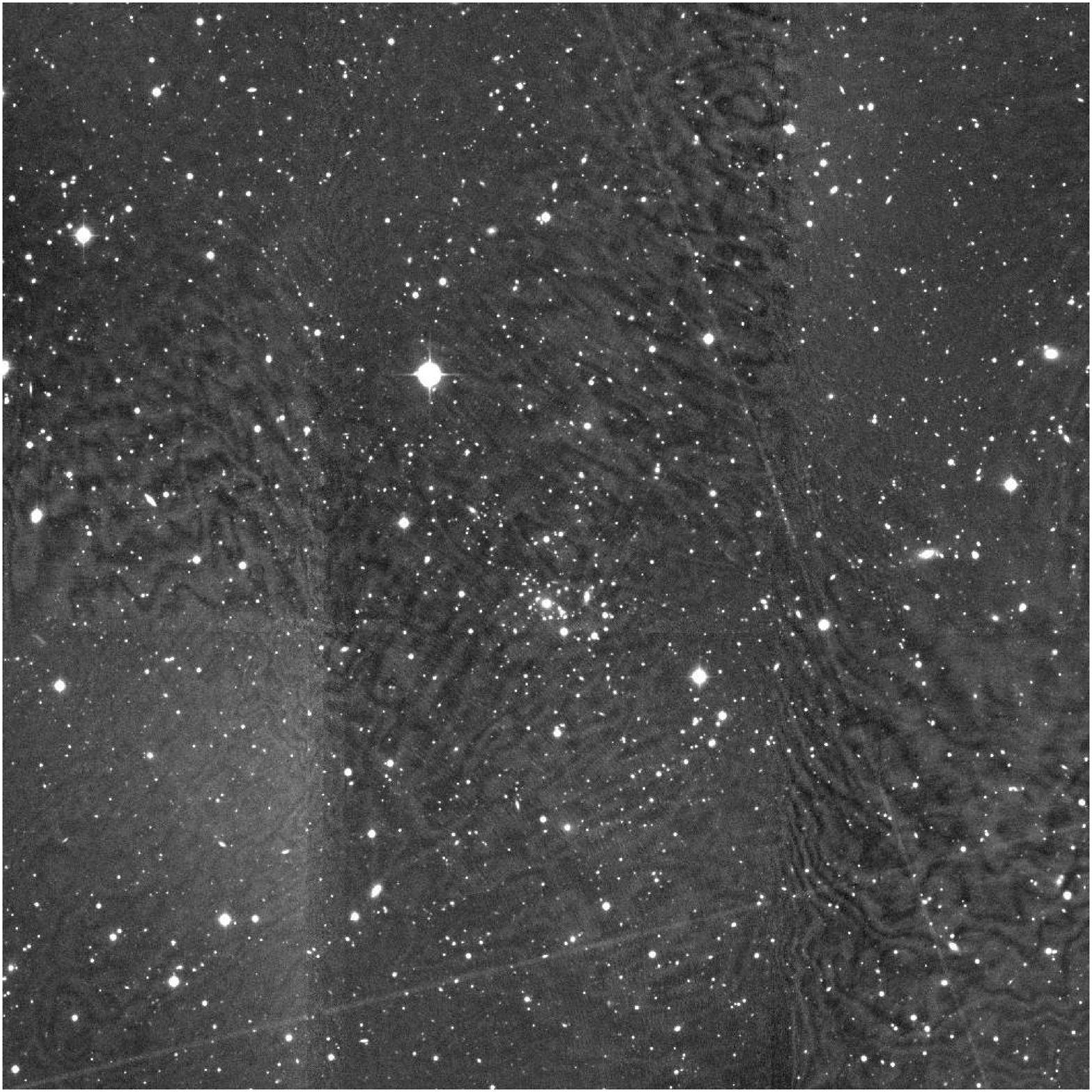}
  \end{minipage}%
\begin{minipage}{.5\textwidth}
  \centering
  \includegraphics[width=.95\linewidth]{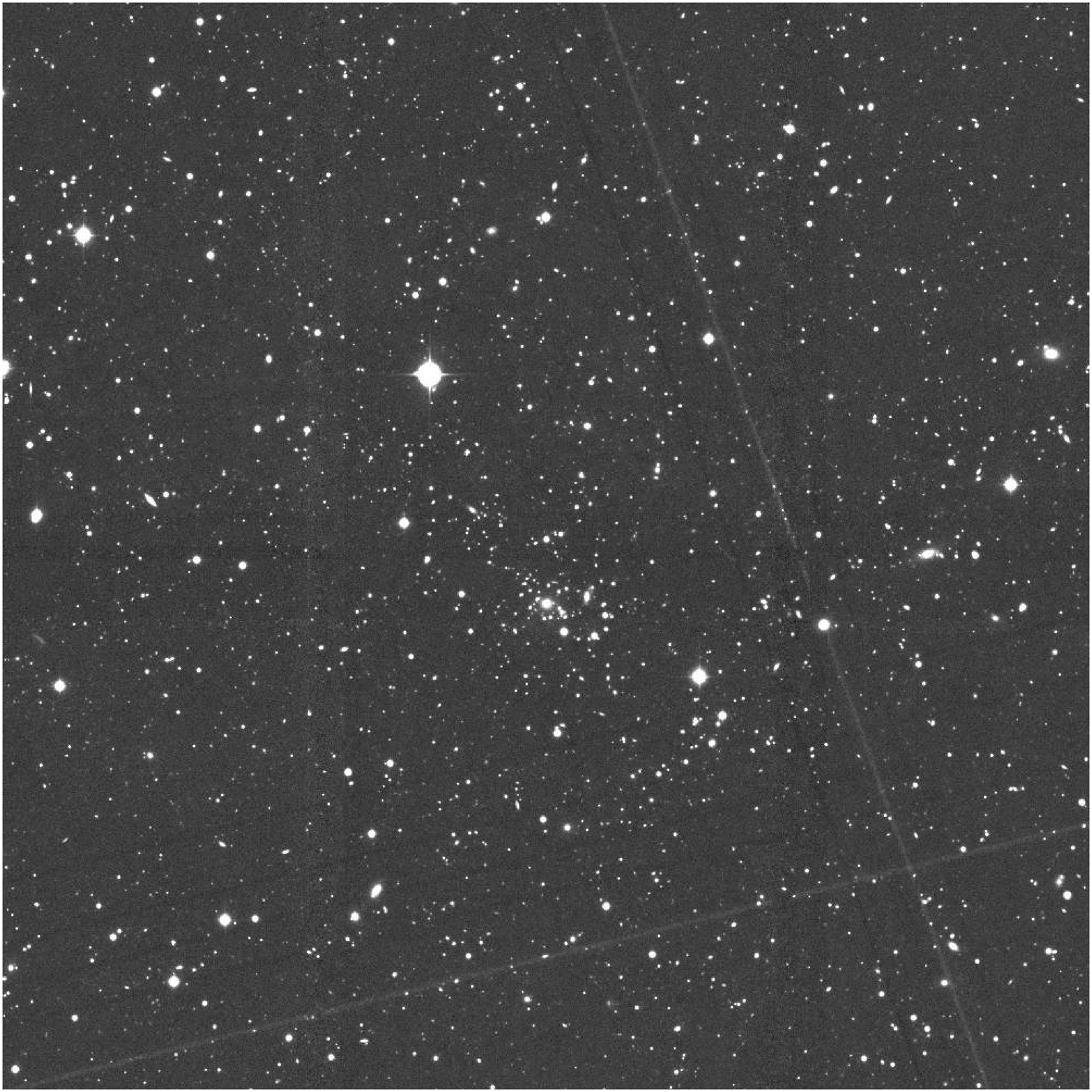}
\end{minipage}
\caption{\footnotesize Illustration of the background residual correction, showing a 15'$\times$15' part of the \texttt{PSZ2 G048.21-65.00} $z$-band stack before (left panel), and after (right panel) the correction. The median improvement in the effective depth of the $z$-band images in a 2$''$ aperture is approximately 0.4 magnitudes.}
\label{fig:backgroundcorr}
\end{figure*}

\subsection{Homogenisation of SZ detections}\label{sec:re-extract}
Given that the PSZ1 catalogue was constructed from the maps of about half the final depth, we re-measure the SZ signal for each candidate consistently in the final maps (including those PSZ1 candidates that did not end up in the PSZ2 catalogue). We perform a consistent blind search of the SZ signal around the selected locations in the final version of the \textit{Planck} maps (column SNRblind in Table~\ref{tab:overview}). We use the Matched Multi-Filter 3 \citep[MMF3,][]{melin06} detection method to search and extract SZ parameters from the \textit{Planck} maps. We note that the MMF3 detection method resamples the \textit{Planck} frequency maps centred on a given location, which may slightly affect the significance with which clusters are blindly detected. 

Eight of the PSZ1 entries we considered in 2013A and 2013B are not part of the PSZ2 catalogue as they fall below the significance of S/N=4.5 in the final \textit{Planck} maps. Their S/N drop to even below 3 (except for one at S/N$\sim$4) which suggests that they may have been noise fluctuations in the PSZ1 catalogue. It is informative to perform all processing steps on the available data for these PSZ1-only detections too, and to compare them to other candidates in the final sample as an extra test of our validation method. We will make these comparisons in Sect.~\ref{sec:yesorno}.

\section{Optical data \& catalogues}\label{sec:data}
To confirm the optical counterparts of these cluster candidates by estimating photometric redshifts and optical richnesses, we observed each through the $r$- and $z$-filters using MegaCam mounted on the Canada-France-Hawaii Telescope (CFHT). These filters cover wavelengths such that we expect to obtain reasonably precise photometric redshift estimates over a redshift baseline up to $z\sim 0.8$. The data are pre-processed using the \textit{Elixir} pipeline \citep{elixir04}. After this standard reduction, there are still residual background patterns due to e.g. scattered light, fringe residuals, and amplifier drift. Given that these patterns are reasonably stable over time, and that most of the data for a given field and filter are taken during the same night, we can correct for these background effects. We do this by using the dithered pattern of observations to differentiate signals that are on a fixed position on the CCD array from sky-bound signals, similar to our approach in \citet{vdB13,vdB15}. An example of this procedure is shown in Fig.~\ref{fig:backgroundcorr}. We remove cosmic rays on a frame-by-frame basis by using the Laplacian Cosmic Ray Identification method \citep{dokkumcosmics}.

Astrometric solutions for the data are obtained using \texttt{SCAMP} \citep{scamp}, based on the USNO-B1 reference catalogue. We combine all exposures taken with a given filter for each semester to effectively increase the source density and obtain highly precise solutions with an internal scatter between filters of $< 0.05''$. Relative photometric zeropoints between exposures are estimated based on overlapping sources between different frames. 

Although the exposures for a given field and filter are generally taken consecutively during the same night and therefore have a similar image quality (IQ), there can be a substantial difference between the image quality of the $r$- and $z$-band images of a given field, see Table~\ref{tab:dataoverview}. To measure colours on the same intrinsic part of the galaxies in both filters, we have to account for these differences. The approach we take is to use \texttt{PSFEx} \citep{psfex} to determine a shapelet-based convolution kernel for each filter, to make the PSF homogeneous between the two stacks. As target PSF we choose a Moffat profile with a FWHM that is 10\% larger than the largest IQ of the $r$- and $z$-band stacks, with a Moffat-$\beta$ parameter of 2.5. These choices ensure that the target PSF has sufficiently broad wings that no de-convolution is required. 

The exposure times of our images are chosen such that we probe the red-sequence galaxies, which dominate the cluster galaxy population, to a similar limit in the $r$- and $z$-band. We choose to use the original, unconvolved $z$-band image for source detection, as it probes the rest-frame galaxy redward of the 4000$\AA$ break for redshifts up to $z\sim 1$, and thus provides catalogues that are closest to being stellar-mass selected. We measure colours using circular apertures with a diameter of 2$''$ on the seeing-homogenised stacks.  At this stage we re-evaluate the photometric zeropoints in several steps. First, for the images that overlap with the SDSS DR9 \citep{sdssdr9} footprint (19 out of 28 fields), we compare the $z$-band MAG\_AUTO magnitude from MegaCam with the model magnitude $z$-band magnitude from SDSS, and adjust our zeropoint to match this reference. We do the same for the $r$-band, based on the difference between the $r-z$ aperture colour measured from the MegaCam data, and the SDSS model $r-z$ colour. While making these comparisons, we apply the linear colour terms between the SDSS and MegaCam filters, as listed on the CFHT website\footnote{http://www4.cadc-ccda.hia-iha.nrc-cnrc.gc.ca/en/megapipe/docs/filt.html}. Typical corrections are on the order of 0.05mag, and are largest in the case of non-photometric observing conditions (which happened mostly in semester 2013A).

We exploit the colour-colour distribution of stars as a second reliable photometric calibrator. To be able to calibrate the data against a universal stellar locus, we require a third photometric band. Because the number of stars that are bright enough in 2MASS, but unsaturated in the deep MegaCam exposures, is limited, we find that the WISE 3.4$\mu$m band serves as a better reference. Therefore we construct an empirical $r-z$ vs $z-$3.4$\mu$m colour-colour diagram of bright stars, combining all fields that suffer from little Galactic dust extinction \citep{schlegel98} that have been calibrated against SDSS. Subsequently, we re-calibrate the remaining fields (9 out of 28 fields) by comparing the measured colours with this $r-z$ vs $z-$3.4$\mu$m reference stellar locus. Again, corrections are on the order of 0.05mag. We estimate the remaining systematic uncertainty on the $r-z$ colour, especially for fields with significant Galactic dust extinction, to be on the order of 0.05. We reach a median 5-$\sigma$ aperture magnitude depth of 25.0 and 23.8 in the $r$- and $z$-band stacks, respectively. Table~\ref{tab:dataoverview} gives an overview of the basic properties of the data per field.

\begin{table*}
\caption{\footnotesize Characteristics of the MegaCam imaging data taken for the 28 fields.}
\begin{center}
\begin{tabular}{lcccccc}
\hline \hline 
Field&$r$-band IQ$^{\mathrm{a}}$&$r_{\mathrm{ lim,2''}}^{\mathrm{b}}$&$z$-band IQ$^{\mathrm{a}}$&$z_{\mathrm{ lim,2''}}^{\mathrm{b}}$&     $z_{\mathrm{ lim,tot}}^{\mathrm{c}}$&Limiting\\
&PSF FWHM ['']&[mag$_\mathrm{AB}$]&PSF FWHM ['']&[mag$_\mathrm{AB}$]&[mag$_\mathrm{AB}$]&Redshift$^{\mathrm{d}}$\\
\hline
\texttt{PSZ1 G023.38-33.46}&1.02&25.03&0.82&24.02&23.3&0.99\\
\texttt{PLCK G027.65-34.27}&0.56&24.97&0.50&23.77&23.7&1.08\\
\texttt{PSZ1 G031.41+28.75}&0.93&25.06&1.08&23.50&22.5&0.76\\
\texttt{PSZ2 G037.67+15.71}&0.83&24.46&0.80&23.30&22.9&0.83\\
\texttt{PSZ1 G038.25-58.36}&0.79&24.94&0.74&23.81&23.2&0.96\\
\texttt{PLCK G038.64-41.15}&0.54&24.82&0.51&23.89&23.8&1.11\\
\texttt{PSZ2 G041.69+21.68}&0.74&24.83&0.87&23.81&23.1&0.92\\
\texttt{PSZ2 G042.32+17.48}&0.54&24.45&0.55&23.22&23.2&0.93\\
\texttt{PSZ2 G048.21-65.00}&0.64&25.10&0.92&23.94&23.1&0.94\\
\texttt{PSZ1 G051.42-26.16}&0.73&25.14&0.87&24.02&23.2&0.94\\
\texttt{PLCK G053.41+61.50}&0.53&24.94&0.58&23.53&23.3&1.00\\
\texttt{PSZ1 G053.50+09.56}&0.51&24.05&0.90&22.74&22.5&0.63\\
\texttt{PSZ2 G071.67-42.76}&0.77&25.30&0.81&24.18&23.5&0.99\\
\texttt{PSZ2 G071.82-56.55}&0.52&24.96&0.50&23.71&23.6&1.07\\
\texttt{PSZ2 G076.18-47.30}&0.64&24.79&0.48&23.66&23.4&1.01\\
\texttt{PLCK G079.95+46.96}&0.53&24.80&0.49&23.68&23.6&1.08\\
\texttt{PSZ1 G081.56+31.03}&0.84&25.00&0.76&23.77&23.2&0.96\\
\texttt{PLCK G087.58-41.63}&0.53&24.83&0.48&23.74&23.6&0.98\\
\texttt{PSZ1 G092.41-37.39}&0.61&24.95&0.59&23.74&23.4&0.91\\
\texttt{PSZ2 G106.15+25.75}&0.69&25.05&0.93&23.69&22.8&0.83\\
\texttt{PSZ2 G119.30-64.68}&0.57&24.77&0.52&23.73&23.5&1.05\\
\texttt{PSZ2 G141.77+14.19}&0.91&25.11&0.75&23.79&23.2&0.85\\
\texttt{PSZ2 G157.07-33.63}&0.60&25.11&0.62&23.81&23.4&0.88\\
\texttt{PLCK G191.75-21.78}&0.68&24.99&0.58&23.81&23.4&0.97\\
\texttt{PSZ2 G198.80-57.57}&0.75&24.78&0.59&23.77&23.3&0.98\\
\texttt{PSZ2 G208.57-44.31}&0.65&24.96&0.69&23.72&23.2&0.96\\
\texttt{PLCK G227.99+38.11}&0.56&25.14&0.62&23.87&23.5&1.04\\
\texttt{PSZ1 G240.42+77.58}&0.66&25.09&0.66&24.11&23.6&1.08\\
\hline
\label{tab:dataoverview}
\end{tabular}
\end{center}
\begin{list}{}{}
\item[$^{\mathrm{a}}$] PSF size of the stack, before homogenisation.
\item[$^{\mathrm{b}}$] 5-$\sigma$ limiting magnitude in a circular aperture with a 2$''$ diameter, after PSF homogenisation.
\item[$^{\mathrm{c}}$] 80\% detection limit estimated from the recovery of small simulated galaxies injected in the $z$-band image.
\item[$^{\mathrm{d}}$] Redshift at which the 80\% limit reaches down to magnitude $m_{z}^{*}+1.00$, accounting for Galactic dust extinction.
\end{list}
\end{table*}

\section{Redshift- \& richness estimates}\label{sec:analysis}
\subsection{Red-sequence model}\label{sec:rsmodel}
Our analysis is based on the properties of red-sequence galaxies, which are highly abundant in galaxy clusters, at least up to $z\sim 1$, and thus provide a signal with a high contrast against the background. To interpret our data, we first construct an empirical model that predicts the colour of red-sequence galaxies as a function of magnitude and redshift. We exploit the 30-band photometric data from the COSMOS/UltraVISTA field \citep{muzzin13a}, from which we select galaxies over a range of redshifts with similar properties as our cluster red-sequence galaxies. By combining the excellent photometric redshifts from this field with U-V and V-J rest-frame colour measurements, we select red-sequence galaxies down to faint magnitudes ($z_\mathrm{tot}\approx 24.0$), in redshift bins up to $z=1.1$ \citep[e.g.][]{williams09,vdB13}. Note that the r$+$ and z$+$ Subaru filters, which have been used in the UltraVISTA catalogue, are significantly different from the $r$- and $z$-band MegaCam filters used in this analysis. To make the model applicable to our data set, we thus match the COSMOS/UltraVISTA catalogue to the CFHTLS D2 field catalogue \citep{erben09,hildebrandt09a}, which overlaps with the COSMOS field. By selecting galaxies from the 30-band catalogue, while using the flux measurements from the CFHTLS catalogue, we obtain a catalogue of quiescent galaxies with MegaCam $r-z$-colours as a function of total $z$-band magnitude and redshift. 

\begin{figure*}
\resizebox{\hsize}{!}{\includegraphics{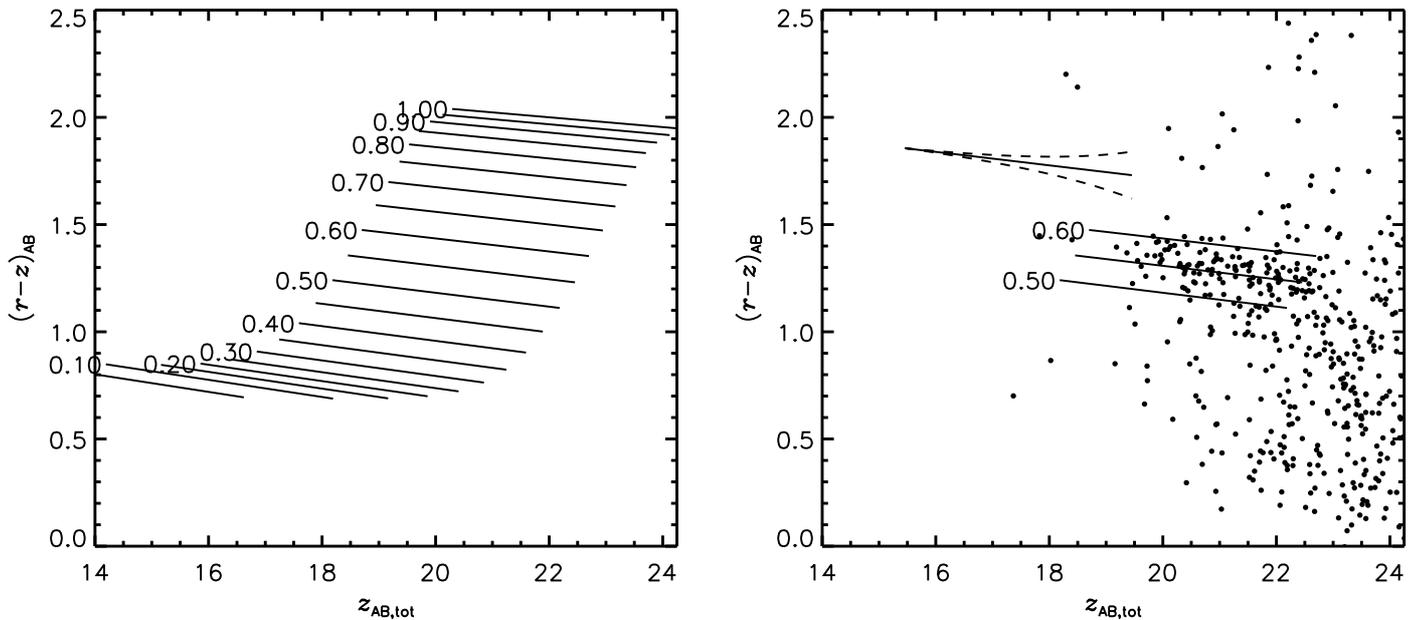}}
\caption{\footnotesize \textbf{Left panel:} \textit{Black lines:} Empirical model of the red sequence. At each redshift, the lines run from $m^*-2<m<m^*+2$. \textbf{Right panel:} Three model lines with galaxy colours and magnitudes overplotted from field \texttt{PSZ2 G119.30-64.68} ($z_\mathrm{spec}=0.557$), within 2 arcmin from the centre of the overdensity. For clarity, error bars are not shown on individual points, but these increase in size from lower-left to upper-right. The inset illustrates the statistical error on the colour measurement around the location of the red sequence at this redshift (offset from the $z=0.55$ model for clarity).}
\label{fig:rsmodel}
\end{figure*}

Next we fit a linear relation to these colours, in overlapping redshift bins with width 0.04 and stepsize of 0.01. We remove outliers, especially with bluer colours (since cluster red-sequence galaxies are expected to be the oldest and thus reddest at a given redshift). For each redshift bin we thus obtain a slope, intersect (at a magnitude of $z_\mathrm{tot}$=22.0 to reduce covariance between estimated slope and intersect), and scatter around the sequence. We subsequently fit a polynomial relation to each of these three parameters as a function of redshift, to obtain smoothly varying functions, which we find to describe the colours of these quiescent galaxies well. The derived red-sequence model in the MegaCam $r$- and $z$-band filters is shown in Fig.~\ref{fig:rsmodel}.
When using this model to estimate redshifts, the highest precision can be obtained in the regime where the 4000$\AA$ break is located between the observed $r$- and $z$-band filters, which is in the redshift range $0.35 \lesssim z \lesssim 0.80$. Outside this range, the $r$- and $z$-bands lose their constraint on the redshift, although the apparent $z$-band magnitude distribution of cluster galaxies may still be used as a rough measure of the distance modulus. 

Besides the choice of filters, the depth of the data also limits the detectability of high-redshift clusters. We estimate 80\% detection completeness limits for the $z$-band stacks, based on the recovery of simulated galaxies which we inject in our images. We assume S\'ersic light profiles with a constant S\'ersic parameter of n=4. We draw sizes from a uniform distribution with effective radii between 1-3kpc (assuming an angular diameter distance corresponding to redshift $z=0.6$), which is appropriate for sources around our detection limit. Note that the recovery of simulated sources is only mildly dependent on these parameter choices, since they are poorly resolved in our ground-based images, resulting in a recovery rate that is primarily dependent on the PSF size. The faintest magnitudes at which 80\% of injected sources are still detected, are shown in Table~\ref{tab:dataoverview}. We define corresponding redshift limits as the redshift at which this magnitude limit reaches down to magnitude $m_{z}^{*}+1.00$. We base our estimate of $m_{z}^*$, the characteristic magnitude in the $z$-band, on the stellar mass functions measured in \citet{muzzin13b,ilbert13}, which suggest that the characteristic mass of quiescent galaxies in our redshift range is approximately described by $M^{*}_\mathrm{star}\approx10.95-0.167\times\rm{Redshift}$. The characteristic $z$-band magnitude we use corresponds to the magnitude of a quiescent galaxy formed at $z_{\mathrm{form}}=3$ that has a stellar mass of $M^{*}_\mathrm{star}$. The conservative limit of $m_{z}^{*}+1.00$ ensures that we can estimate richnesses without depending too much on an extrapolation of the luminosity function below the detection limit (see Sect.~\ref{sec:richnessest}).

We perform an automated search for red-sequence galaxies in the colour-magnitude diagram (de-reddened for Galactic dust) as a function of redshift, by comparing the observed $r-z$ colour of galaxies with this empirical red-sequence model. For each redshift from $z$=0.05 to the limiting redshift per field, with d$z$=0.01, we create a map of galaxies with $r-z$ colours that are consistent with this model, allowing for an increase in photometric scatter towards the faint end. This does not (yet) provide a complete census of the galaxy population in these systems, but these maps contain a near-optimal signal for an overdensity of red-sequence galaxies at a given redshift \citep[e.g.][]{gladdersyee00}. We then consider as a possible centre of the galaxy overdensity those galaxies (independent of colour) which are (1) located within 4$'$ from the SZ detection, and (2) brighter than $m<m^*$ at this redshift.  Around each of these possible centres we count the number of possible red-sequence galaxies within a radius of 0.5 Mpc, and perform a statistical background subtraction by performing the same colour selection on the regions around the overdensities. We then select, for each field, the location of the most significant overdensity. These coordinates are listed in Table~\ref{tab:overview}. We verify that these centres are generally close to the locations of visually confirmed galaxy overdensities.

\subsection{Photometric redshift estimates}
\begin{figure}
\resizebox{\hsize}{!}{\includegraphics{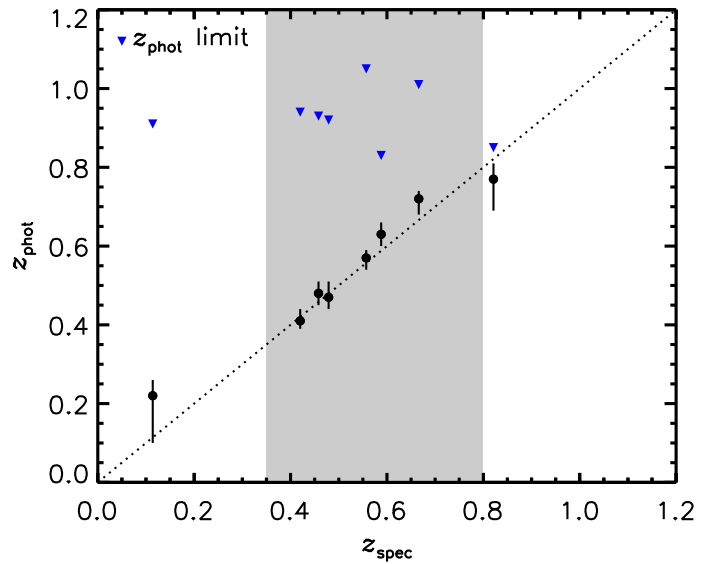}}
\caption{\footnotesize Spectroscopic versus photometric redshifts for the systems that have been confirmed spectroscopically. The grey area highlights the redshift-interval within which we can estimate photometric redshifts most precisely, due to our choice for the $r$- and $z$-bands. Blue triangles mark the approximate redshift depth of the data (cf. Sect~\ref{sec:rsmodel} and Table~\ref{tab:dataoverview}).}
\label{fig:speczphotz}
\end{figure}

The method described above is inadequate to measure precise redshifts of the galaxy overdensities. The exact vertical location of the red sequence (which best constrains the photometric redshift) is washed out by the relatively large width of the search box (which was chosen to optimise the signal of the detection). To improve the redshift estimate, we repeat the above procedure, but fix the location and perform a search in a narrower colour-box to specifically determine the location of the red sequence. We use a box with a fixed width of 0.05, which roughly equals the systematic uncertainty left in our $r-z$ colour calibration. Our best redshift estimate is the one that provides the model that maximizes the number of galaxies in the box around it. In Table~\ref{tab:overview} we provide these values, together with a 68\% error estimate. This uncertainty interval corresponds to redshift values for which the number drops by less than 1$\sigma$ compared to the number of galaxies in the box corresponding to the best redshift. For the clusters that have a spectroscopically confirmed redshift, we find overall consistency, within the uncertainties, between these redshifts and our photometric estimates, as illustrated in Fig.~\ref{fig:speczphotz}. 

\subsection{Richness estimates}\label{sec:richnessest}
In both methods described above, the used selection box is too small to account for all galaxies that appear to be offset from the red-sequence model due to photometric (and intrinsic) scatter. Since this renders these galaxy numbers inaccurate, we perform a third and final analysis in which we fix the location and redshift of the model, and expand the width of the selection box around the model to obtain a more complete sampling of red-sequence galaxies that are associated with the cluster. As we do this, the statistical background correction becomes more uncertain and imprints a larger component to the overall error on the richness estimate. We expand the box until it has a width of two times the estimated (intrinsic+statistical) scatter of galaxies around the red sequence. We make a small correction to account for galaxies with a larger scatter, which is expected to be $\sim$5\% of the total, assuming Gaussian scatter.

To be able to compare these results to the mass-richness relation of \citet{rykoff14,rozo15}, we make our richness measure comparable to the richness estimator $\lambda$ used in those studies. Therefore we (1) consider galaxies with magnitudes brighter than $m<m^*+1.75$, and (2) make the radius ($R_{c}$) deviate from 0.5 Mpc and increase it until 
\begin{equation}
n=100\left( \frac{R_{c}}{R_{0}} \right)^{1/\beta},
\end{equation}
where $R_{0}=1.0 h^{-1}$ Mpc, and $\beta=0.2$, following Equation 4 in \citet{rykoff14}. Note that, although we do not assume a radial profile for the galaxy population, in some cases we have to extrapolate the richness measurement from the detection limit to $m=m^*+1.75$. Measurements of the luminosity function of cluster galaxies have indicated that the slope of the distribution is quite shallow up to that magnitude limit \citep[$\alpha\approx -1.2$, e.g.][]{barkhouse07,moretti15}. Completeness correction factors are therefore generally small, and only mildly dependent on the exact slope of the luminosity function. Two notable exceptions are \texttt{PLCK G087.58-41.63} and \texttt{PSZ2 G141.77+14.19}, for which we have to correct the richness for incompleteness using correction factors of 2.1 and 1.8, respectively. 

The richnesses are listed in Table~\ref{tab:overview}. The associated uncertainties we give are purely statistical; they are the quadratic sum of the Poisson errors on the pure cluster+background counts, and Poisson errors on the subtracted background. It does not include a propagation of the redshift uncertainty on the richness measurement, nor the uncertainty on the correction factor (which we applied in 7 of the 28 fields). The statistical uncertainty we account for dominates over the other sources of uncertainty, except for two systems which we later confirm as clusters: \texttt{PSZ2 G071.82-56.55}, which has a relatively small statistical uncertainty on the richness measurement, but a relatively uncertain (high) redshift, and \texttt{PSZ2 G141.77+14.19}, which also has a relatively uncertain high redshift and a large correction factor. However, note that including the systematic uncertainty would not have an effect on this analysis, as it does not change the sample of clusters we confirm in Sect.~\ref{sec:criteria}.

\begin{figure*}
\resizebox{\hsize}{!}{\includegraphics{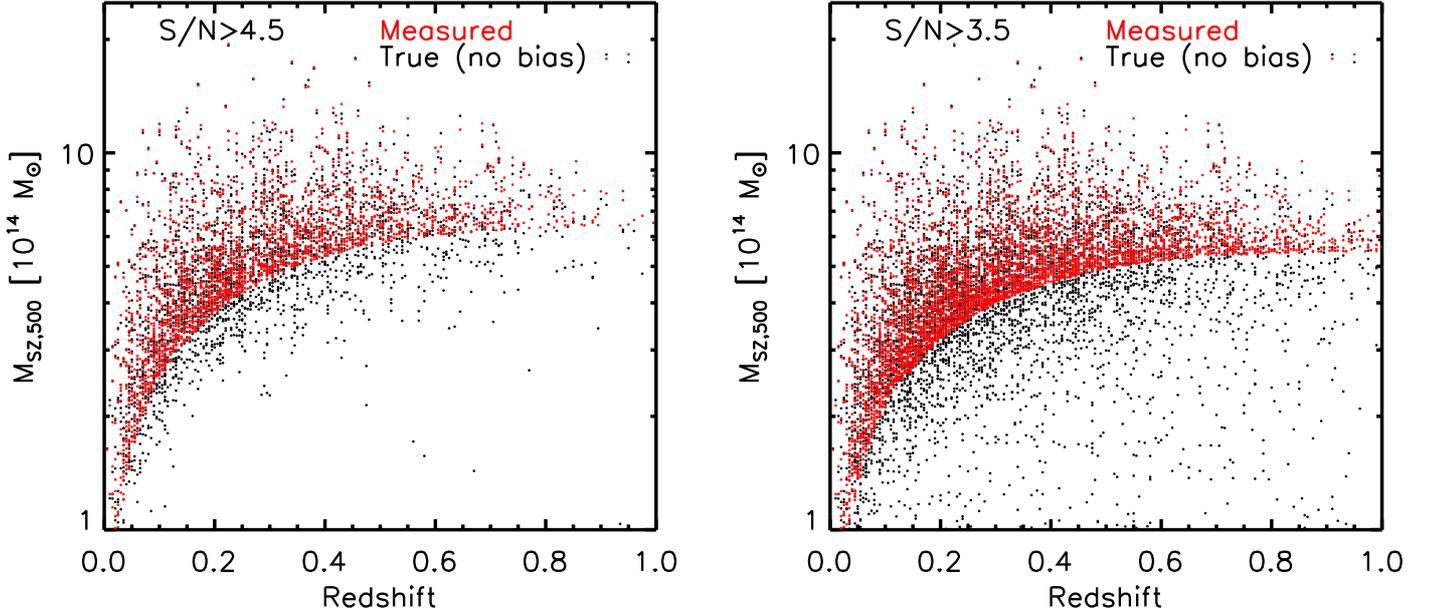}}
\caption{\footnotesize Illustrating the effect of Eddington bias on the measured SZ-based masses, after a significance cut of S/N$>$4.5 (left) or S/N$>$3.5 (right) is applied. Red: Measured mass distribution of sources as a function of mass and redshift in 30,000 sqr degrees. Black: True mass distribution of the same systems.}
\label{fig:eddington}
\end{figure*}

\section{Likelihood of counterparts}\label{sec:counterparts}
Galaxy overdensities appear on a range of different scales, from low-mass groups to massive galaxy clusters. Even if spectroscopic information indicates that a system of galaxies is physically associated, it is important to define quantitative criteria to assess whether a given system is consistent with what we expect for a halo that is responsible for the measured SZ signal. The measured SZ signal provides a halo mass estimate through the procedure introduced in Arnaud et al. (in prep.), as was already applied in \citet{psz1,psz2} to estimate masses for clusters with confirmed redshifts in the PSZ1 and PSZ2 catalogues, respectively. Before we exploit the relationship between richness and halo mass, we study how noise in the \textit{Planck} maps affects the SZ-based halo mass proxy, especially in the low significance regime at which we are detecting clusters.

\subsection{Eddington bias in the SZ halo mass proxy}\label{sec:eddington}
Due to the presence of noise in the \textit{Planck} maps, relatively low-mass haloes may scatter over the SZ-significance threshold and make it into the catalogue. Due to the steepness of the halo mass function \citep[e.g.][]{tinker08,bocquet15}, it is expected that more low-mass haloes scatter upward than high-mass haloes scatter downward. This results in a distorted view of the underlying halo distribution after applying a cut in significance (\citealt[][]{eddington1913} bias). Note that this bias is purely statistical \citep[also see e.g. Appendix A in][for an illustration of the effect of this type of statistical bias]{mantz10}. The cosmological analysis based on \textit{Planck} cluster number counts restricts itself to the most significant SZ detections (S/N$>$6), and the Eddington bias is moderate for this high S/N cut \citep[e.g. Appendix A.2 of][]{planckclustercosmo14}. However, we consider SZ detections down to S/N=4 and even below, which brings us to the regime where this type of bias starts to play a significant role.

A full accounting of the effect of Eddington bias on our analysis would require us to insert modelled SZ profiles in simulated \textit{Planck} maps with representative noise properties, and we leave this to a future study. We rather provide the following estimate of the effect, in which we assume that haloes are spatially independent (i.e. non-overlapping). We simulate a list with masses and redshifts of all haloes with mass $M_{500} > 10^{14}\, \mathrm{M_{\odot}}$ up to a redshift of $z=1.25$ in a representative lightcone that spans 30,000 square degrees on the sky. For this we follow the \citet{tinker08} halo mass function and the redshift-dependent comoving volume element for our assumed cosmology. The next step is to estimate, given the noise properties of the \textit{Planck} maps, at what significance a source with a given $M_{500}$ and redshift would be detected. For this we first use Equation 7 \& 8 in \citet{planck15clustercosmology} to relate these masses and redshifts to a $Y_{500}$ and $\theta_{500}$. A hydrostatic mass bias of 1-b=0.8 is assumed here, which is the baseline value used in \citet{planck15clustercosmology}, and is supported by e.g. a weak-lensing study of \citet{hoekstra15}. We take, from the \textit{Planck} noise maps, the average noise value $\sigma_{Y_{500}}$ over the SZ catalogue region \citep[i.e. the final version of Fig.~4 in][]{planckclustercosmo14}. This noise value depends on the aperture considered, $\theta_{500}$, and was shown to be approximated by a Gaussian distribution \citep[Sect.~3.3 in][]{planck15clustercosmology}. By combining $Y_{500}$ and the appropriate noise value, we obtain a significance for each halo. When we compare relations between mass and significance, for a given redshift, we find that these are in excellent agreement with values of \texttt{SNR} and \texttt{M500} in the published PSZ2 catalogue, as they should be.

In the presence of noise, this ``true'' significance deviates from the measured significance, which we model by adding a random variable drawn from a standard normal distribution to the ``true'' significance. If we apply a cut to the measured significance of S/N$>$4.5, we obtain a total of 1359 sources. Given that the PSZ2 catalogue used this significance threshold, it is reassuring that this number is comparable to the number of detections reported in PSZ2 (1653), and has a roughly similar redshift distribution. After converting this measured significance back into a measured mass, following the same equations as before, we obtain the red points in Fig.~\ref{fig:eddington}. There is a sharp line below which no clusters are detected, which results from the direct relationship between significance and estimated mass. In the real data this sharp edge is slightly diluted, because the noise properties are not completely isotropic, and in some studies the SZ signal is re-measured at the location of an optical overdensity, which slightly reduces the SZ-based mass at fixed blind significance. When we compare these measured masses to the true masses of the same haloes (black points in Fig.~\ref{fig:eddington}), the nature of the Eddington bias becomes apparent. If we lower the significance cut from 4.5 to 3.5, the bias becomes more severe, as seen in the right-hand panel.

\begin{figure}
\resizebox{\hsize}{!}{\includegraphics{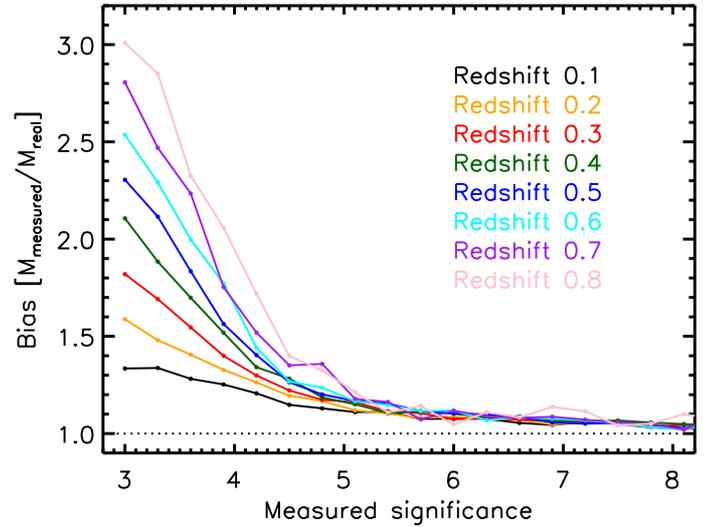}}
\caption{\footnotesize Magnitude of Eddington bias in the \textit{Planck} SZ mass proxy as a function of measured significance and redshift. Shown is the mean fractional difference between the measured mass and the true mass. Since we only consider haloes in the mass range $M_{500} > 10^{14}\, \mathrm{M_{\odot}}$, this is a lower limit to the true bias. Given this skewed distribution, the mean is higher than the median by up to $\sim$50\%.}
\label{fig:biasplot}
\end{figure}

In Figure~\ref{fig:biasplot} we quantify the magnitude of the bias as a function of measured significance. The magnitude of the bias depends on the steepness of the halo mass function around a given significance, and is thus redshift-dependent. For this plot we repeat the experiment 10 times, and thus consider 300,000 square degrees to improve the statistics on these numbers. Given the nature of this effect, it would be useful to estimate masses in a way that is not affected by Eddington bias. This illustrates the necessity of deeper follow-up data in cases of detections near the survey limit. Although mass proxies based on optical or X-ray follow-up data are considered to be less accurate than the ones that are SZ-based, these provide measurements that are independent of the detection, and are thus \textit{not} subject to the bias. 

\subsection{Mass-Richness relation}\label{sec:massrichness}
One way to verify our candidates is to compare the measured richnesses to the SZ-based halo mass proxy, as was done in \citet{rozo15,psz2}. Note that we searched within a radius of 4$'$ for the most significant galaxy overdensities around each SZ peak. For a galaxy overdensity that is found far away from the SZ maximum, the SZ signal at that location may be significantly lower. Before we estimate the SZ-based mass, we therefore re-extract the SZ signal at the location of the galaxy overdensity. This decreases the significance compared to the significance of the blind detection (by definition, see Table~\ref{tab:overview}), where the given distance is an integer number of pixels (1 pixel = 1.72$'$) on this 2-dimensional grid between the blind detection and the optical centre. This distance and the difference in S/N of the blind detection and the re-extracted value may also serve as a check on the identified counterpart. Note however that some of the clusters in our sample are multi-modal in their galaxy distribution, as also suggested by the figures in Appendix~\ref{sec:images}, where white circles mark the centres of the assumed optical position. 

\begin{figure}
\resizebox{\hsize}{!}{\includegraphics{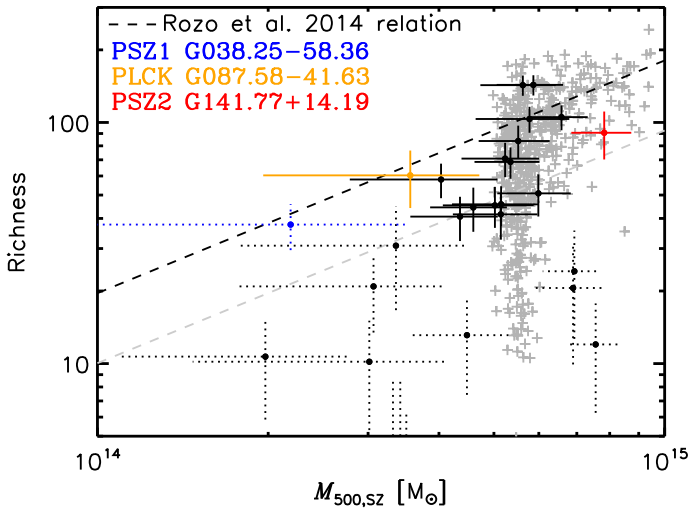}}
\caption{\footnotesize Comparison between \textit{Planck} SZ halo mass proxy, and richness (similar to the definition of \citet{rykoff14}. The mass-richness relation from \citet{rozo15} is overplotted. The grey '+'-symbols show the distribution of a simulated sample with redshift $z>0.5$ and measured significance S/N$>$3.5, and is thus similar to the population from which our candidates have been selected. Both intrinsic and statistical scatter in the richness have been included, but most of the scatter in the grey distribution is caused by statistical uncertainties in the SZ mass proxy. The effect of Eddington bias is apparent, since it causes the observed points to deviate from the intrinsic relation. Our criteria to confirm cluster counterparts (solid error bars) are described in Sect.~\ref{sec:criteria}, and are largely based on the grey dashed line. Three \textit{Planck} detections are highlighted in colour, and these are discussed in particular in Sect.~\ref{sec:candidates}.}
\label{fig:massrichness}
\end{figure}

Figure~\ref{fig:massrichness} shows the resulting comparison between mass and richness. The black dashed line is the best-fitting relation from \citet{rozo15}, which is based on a comparison between \textit{Planck} and the redMaPPer cluster catalogue. \citet{rozo15} estimated an intrinsic scatter of $\sim$25\% around this relation. The low-mass end of the relation is constrained using haloes at low redshift, which are thus still significantly detected in the \textit{Planck} maps. This assumes that the richness-mass relation of galaxy clusters does not evolve over this redshift range \citep[e.g.][]{andreoncongdon}. Note however that at higher redshift ($z\gtrsim1.0$), an increasingly large population of galaxies may not yet be part of the red sequence, which complicates the use of a richness-based mass proxy.  

When we compare the mass and associated richnesses for our candidates with the reference relation, the effect of Eddington bias is apparent. The SZ-based mass is likely to over-estimate the true mass, which is probed in an unbiased way by the richness. This can be illustrated further using the simulations detailed  in Sect.~\ref{sec:eddington}. We apply a cut on the measured significance of S/N$>$3.5, and consider the redshift range $z>0.5$, which yields 511 detections in our simplified simulation. Assuming that the richness is a proxy of the true mass, though with a 25\% intrinsic scatter for a given mass, and Poisson uncertainties on the richness measurement, these 511 systems are shown by the grey points in Fig.~\ref{fig:massrichness}. The richnesses measured for the targeted candidates are thus roughly consistent with what we expect in the presence of Eddington bias. This illustrates again that, when approaching the detection limit of \textit{Planck}, deeper auxiliary data (be it X-ray, SZ or optical) are required to obtain an accurate mass estimate. In the present case, the richness thus serves as a more accurate estimator of the halo mass than the SZ mass proxy from the survey data, even though it suffers from a $\sim$25\% intrinsic scatter. 

\subsection{Quantifying criteria for cluster confirmation}\label{sec:criteria}
Due to the role noise fluctuations in the \textit{Planck} maps play in defining a cluster sample at low SZ significance, it is a priori unclear where to draw a line between confirmation and invalidation of an SZ cluster candidate with optical data. A reasonable criterion is to require that a measured SZ signal is dominated by the inverse Compton scattering of CMB photons of a halo, rather than by noise in the \textit{Planck} maps. We can rephrase this by requiring that the richness-based halo mass is more than 50\% of the SZ-based mass, or that candidates should lie above the grey dashed line in Fig.~\ref{fig:massrichness}. 

\begin{figure}
\resizebox{\hsize}{!}{\includegraphics{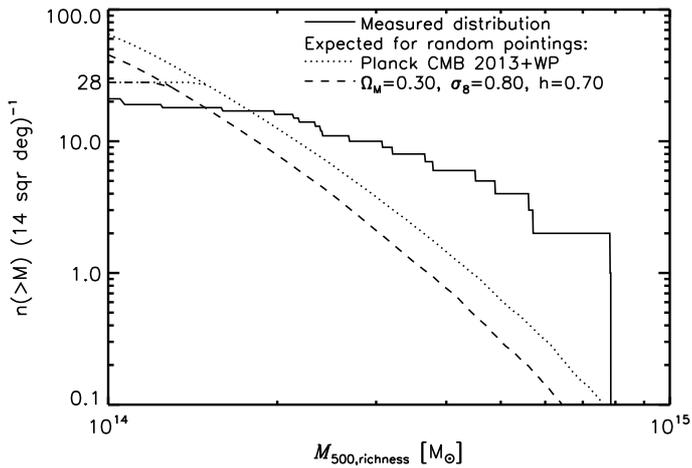}}
\caption{\footnotesize \textit{Solid:} The cumulative distribution of richness-based halo masses measured for our sample of 28 SZ detections, selected from a total area of 14 sqr degree. \textit{Dashed:} Expected cumulative distribution of the richness-based halo mass of the 28 richest systems between $0.4<z<1.0$ per a random area of 14 square degrees, in our assumed cosmology. \textit{Dotted:} Same but for a \textit{Planck} CMB cosmology with WMAP polarisation \citep{planck13cosmology}. A 25\% lognormal scatter on the richness at fixed halo mass is assumed, and each curve includes hydrostatic mass bias.}
\label{fig:expectations}
\end{figure}

This criterion alone, however, is not necessarily sufficient, because we have specifically selected our 28 candidates based on a visual inspection of $\sim$1000 WISE and DSS images. These $\sim$1000 locations are coincident with SZ detections, but even if they were completely randomly spread on the sky, some would fall -by chance- on galaxy overdensities. We considered overdensities in WISE (and corresponding non-detections in DSS) within radii of 4$'$ from $\sim$1000 SZ detections. This corresponds to a total considered area of $\sim$14 square degrees. As a very conservative comparison, we estimate the richness distribution of the 28 richest systems that are expected in a randomly chosen 14 square degrees. For this we use the same simulation as before, but select haloes in the redshift range $0.4<z<1.0$, and add 25\% intrinsic scatter on the richness at a given mass. The results are the dashed and dotted lines in Fig.~\ref{fig:expectations}. The solid line shows the measured cumulative richness distribution of the 28 candidates studied here, where we used the mass-richness relation from \citet{rozo15} to obtain a mass for a given richness. Even though we only followed-up 3\% of the $\sim$1000 candidates, we find that the measured cumulative distribution is already in significant excess of the expected distributions for masses $M_\mathrm{richness}>2\cdot10^{14}\,\mathrm{M}_{\odot}$, corresponding to a richness estimator of $\lambda \gtrsim 40$. For systems with a richness-based mass in excess of this limit, we are therefore confident that these are likely part of our sample because of their SZ signal, and not just because they are a rich system coincident with a pure noise peak. It is important to note that this is a very conservative comparison, since only 3\% of the $\sim$1000 SZ detections have been chosen for the present follow-up with deep MegaCam imaging. A full follow-up of all candidates, which would allow for a more realistic comparison, is likely to raise the measured cumulative distribution substantially, especially around richness-masses of $\sim$2-4$\times10^{14}\,\mathrm{M}_{\odot}$. Since a full estimate of the sample completeness is beyond the scope of this work, we choose to follow this rather conservative limit with this word of caution.

In summary, we require successful candidates to \textbf{(1) have a richness estimate in excess of $\lambda \geqslant$40, and (2) have a richness-based halo mass estimate that is, within 1-$\sigma$, more than 50\% of the SZ-based mass}. Cluster candidates that fail to meet one of these criteria are shown with a dotted error bar in Fig.~\ref{fig:massrichness}. Three candidates are marked in colour, and these are discussed in particular in Sect.~\ref{sec:candidates}. First we discuss some general characteristics of the sample of \textit{Planck} clusters which were confirmed based on the two main criteria, which are shown with solid error bars in Fig.~\ref{fig:massrichness}, and which are listed in the upper part of Table~\ref{tab:overview}. 

\subsection{Remarks on confirmed clusters}\label{sec:yesorno}
To confirm \textit{Planck} cluster candidates we exploit only two main criteria, which are primarily based on the measured richness (see above). There is more information available on these candidates, some of which is also compiled in Table~\ref{tab:overview}. Based on this, the following remarks can be made: 
\begin{itemize}
\item Out of the 16 \textit{Planck} detections that were listed in the PSZ1 catalogue, eight were included also in PSZ2. Each of the five PSZ1 sources that we confirm to be associated with a cluster is included also in the PSZ2, whereas most PSZ1 sources that turn out to be false candidates (8/11) dropped out of the PSZ2 catalogue. This is expected from noise arguments when the depth of the data increases. 
\item Seven (nine) of the 16 PSZ1 sources are listed as \texttt{CLASS1} (\texttt{CLASS2}) in the catalogue. Each of the five confirmed PSZ1 sources was listed as \texttt{CLASS1}, which illustrates the value of such a classification in cleaning the sample a priori. 
\item The 13 candidates that are present in the PSZ2 catalogue have an SZ quality described by the \texttt{Q\_NEURAL} parameter; see \citet{aghanim15} for details, and Table~\ref{tab:overview} for their values. All 10/13 confirmed have a \texttt{Q\_NEURAL}$>$0.8, whereas 3/13 invalidated candidates have a \texttt{Q\_NEURAL}$<$0.1. This illustrates the potential to use this neural network classification to pre-select the most promising SZ candidates for follow-up studies. This classification is less diagnostic in the case of a noise-dominated \textit{Planck} SED, as illustrated by the \texttt{QN} parameter in the updated PSZ1 catalogue \citep{planckpsz1v215}. From the eight SZ detections in PSZ1 that dropped out of PSZ2, only two have a classification of \texttt{QN}$<0.78$, while none are confirmed to correspond to real clusters.
\item The richness peaks which turn out to be associated with false candidates are found roughly uniformly spread around the SZ detections in the 4$'$ search-disk. In particular, the distance between the blind SZ detections and the optical positions (``Distance'' in Table~\ref{tab:overview}) of confirmed candidates is typically smaller than for invalidated candidates. 
\item Candidates that were not part of PSZ1 (and thus were observed with MegaCam in 2014B), have a higher probability of corresponding to significant clusters. This illustrates the benefit of strengthening our WISE+DSS selection criteria to select the most promising candidates, thus keeping the purity relatively high at low SZ detection significance. 
\end{itemize}

We note that there are several cases where we find a significant overdensity of galaxies, but their richness still indicates that their intrinsic SZ signal is likely too small to contribute significantly to the SZ detection. We discuss some individual candidates in Sect.~\ref{sec:candidates}.

\subsection{Representativity and completeness}\label{sec:complet}
 By pre-selecting candidates optically, we are able to preferentially target clusters in the interesting $z\gtrsim 0.5$ range, while increasing the purity of the sample. The price to pay is an a priori more complex selection function. The low significance of the SZ detection is not an issue per se. That is because the completeness of the catalogue, i.e. the fraction of detected clusters as a function of detection threshold and cluster true observables, can be quantified with Monte Carlo simulations, in which clusters are injected in real maps, as well as from analytical assumptions. These are done extensively for the published catalogues \citep[]{psz1,psz2}. 
 
Note, however, that the pre-selection of candidates provides a catalogue that is not purely SZ-selected. True clusters may be rejected by the pre-selection, for example due to uncertainties in the estimated redshift. 
The exact quantification of this additional incompleteness requires sophisticated modelling of the optical properties of the clusters, and this may be difficult. It will introduce a systematic uncertainty in the catalogue selection function, which may hamper the use of such a cluster sample for precision cosmology.  However, for statistical studies of the cluster population, as probe of structure formation, the key requirement is the {\it representativity} of the sample, i.e. that the optical pre-selection does not introduce specific biases as compared to the parent SZ-selected sample.  This is a much less stringent constraint, which can also be studied a posteriori using multi-wavelength data. Therefore, with \textit{Planck} being the only all-sky SZ survey for the foreseeable future, such an approach may be the only way towards a \textit{sizeable}, yet representative, sample of the most massive clusters at $z\gtrsim 0.5$.

\section{Notes on individual candidates}\label{sec:candidates}
We present our candidates in Table~\ref{tab:overview}, and show colour images, mostly of confirmed candidates, in Appendix~\ref{sec:images}. Below we discuss some special cases of confirmed and invalidated candidates, in turn. 

\subsection{Confirmed}
\texttt{PSZ2 G042.32+17.48} was also confirmed by \citet{planckcanary15}, who find an optical counterpart around this position of $z_\mathrm{spec}=0.458$ (3 galaxies). We confirm this overdensity based on a richness estimate, which is consistent with what we expect for a massive galaxy cluster. Our photometric redshift is fully consistent with the reported spectroscopic redshift.

\texttt{PLCK G087.58-41.63}. Galaxy cluster candidate, potentially at high redshift ($z\gtrsim1$), but the $r-z$ colour loses its constraint on the redshift in that regime. Deeper follow-up data in the Near-IR are required to estimate the redshift and make a better richness estimate. We had to correct the richness for incompleteness by a factor of 2.1.

\texttt{PSZ2 G141.77+14.19} shows a significant overdensity close in position to the centre of the SZ detection. After fixing the redshift to the spectroscopic value ($z_\mathrm{spec}=0.821$), we had to correct the richness for incompleteness by a factor of $1.8$, in part because of a substantial dust extinction in the field ($E(B-V)=0.30$).

\texttt{PLCK G191.75-21.78} shows a significant galaxy overdensity, with a $z\sim 0.6$ extended structure that spans several Mpc on the sky.

\subsection{Invalidated}
\texttt{PSZ2 G037.67+15.71} is part of PSZ1 and PSZ2, with a highly significant (S/N=6.61) SZ detection in the final maps. It was classified as \texttt{CLASS1} source in the PSZ1, but no significant galaxy overdensity is found. A larger search radius around the peak of the SZ detection also does not result in a significant overdensity. A closer inspection of the individual frequency maps indicates the presence of an infrared source, which, given that it is located only 15\degr ~from the Galactic plane, might well be related to Galactic dust. The \texttt{Q\_NEURAL} parameter indicates, along the same lines, that the SED has a different shape than expected for a pure SZ signal. 

\texttt{PSZ1 G038.25-58.36}. A significant overdensity of galaxies found. However, the source dropped out of the PSZ2 catalogue, and after re-extraction the SZ-based mass is consistent with zero. This source has likely ended up in the PSZ1 catalogue by chance.

\texttt{PSZ2 G071.67-42.76} is another example illustrating the possible contribution of other emission mechanisms than the SZ decrement to the measured \textit{Planck} SED, just like \texttt{PSZ2 G037.67+15.71}. This source is a strong emitter at 353GHz, which boosts the significance of the SZ signal (=8.37, cf. Table~\ref{tab:overview}). The \texttt{Q\_NEURAL} parameter of this source indicates that this is a likely false cluster candidate. Indeed we do not find a significant overdensity of galaxies around this location. Both this source and \texttt{PSZ2 G037.67+15.71} illustrate the fact that the Planck catalogue is not $100\%$ pure, even at high S/N \citep[this is expected, see e.g. Fig.~11 in][]{psz2}. Note also that the purity is lower in regions of strong dust emission (the dust mask), where these two candidates are located.

\texttt{PSZ1 G092.41-37.39}. Around this SZ detection, \citet{planckcanary15} found a galaxy overdensity of three galaxies around $z_\mathrm{spec}=0.114$. We also find a mild galaxy overdensity, centred on the same position, with a redshift consistent with theirs (though the $r$- and $z$-band filters give a poor precision at such low-$z$). We expect ROSAT, which is relatively deep at this position, to probe down to lower masses than \textit{Planck} at this redshift; however, no significant source is detected in the RASS maps. We measure an X-ray luminosity at this position in the ROSAT [0.1-2.4] keV band, assuming a redshift of $z=0.114$, of $L_{X} = 5.7\pm4.9\times 10^{42}\, \mathrm{ergs\,s^{-1}}$. Using the $L_{X} - M_{500}$ relation from \citet{pratt09}, we estimate a 1$\sigma$ upper limit to the mass of $M_{500}<5\times10^{13}\mathrm{M_{\odot}}$. Given that this is significantly lower than the mass detection threshold of \textit{Planck} at this redshift, we conclude that this is likely to be a noise fluctuation in the PSZ1 catalogue. Note that the source is detected at lower SZ significance in the final maps (S/N=2.43).

\section{Summary \& conclusions}\label{sec:summary}
This paper presents a detailed analysis of deep $r$- and $z$-band follow-up imaging of 28 SZ cluster candidates detected in the \textit{Planck} maps. The candidates were selected to be likely at $z>0.5$, based on external survey data from DSS and WISE. The follow-up imaging data allows us to search for overdensities of red-sequence galaxies around the SZ detections, estimate precise photometric redshifts over a redshift range of $0.35 \lesssim z \lesssim 0.80$, and to measure richnesses for the overdensities. The richness measurement is an important step in the validation process, since it allows us to assess quantitatively if a system is massive/rich enough as would be expected given the strength of its SZ decrement. As such, it provides a means to also re-evaluate confirmed candidates in the literature, even if these are supported by spectroscopic redshifts. 

Given that we consider sources down to an SZ detection significance of S/N$\sim$3 and even below, the SZ-based mass proxy is particularly subject to Eddington bias. In this regime of low-S/N \textit{Planck} detections, the richness provides a mass estimate that is independent of the SZ mass proxy, and thus not affected by Eddington bias. 

We define quantitative criteria based on which we validate the optical counterparts of SZ detections. These criteria are based primarily on a comparison between the SZ-based mass and the richness-based mass. In order to confirm the optical identification of an SZ counterpart, we (1) require that a measured SZ signal is dominated by the inverse Compton scattering of CMB photons of a halo, rather than by noise in the \textit{Planck} maps, and (2) require that the richness is sufficiently high that it is unlikely that the galaxy overdensity is overlapping -by chance- with an SZ source. Following these criteria, we confirm 16 galaxy clusters to be likely counterparts to SZ detections, 13 (6) having an estimated redshift of $z>0.5$ ($z>0.7$). Their richnesses indicate masses that are typically $2\times10^{14}\lesssim M_{500}/\mathrm{M_{\odot}} \lesssim 10^{15}$. 

This work illustrates the potential of the \textit{Planck} maps to provide SZ samples of the most massive galaxy clusters at high redshift ($z\gtrsim0.5$), selected from the whole sky. It shows that the approach we have adopted, to pre-select candidates based on WISE+(S)DSS, indeed increases the efficiency with which we can construct samples of representative clusters in this redshift regime. It demonstrates the need for deep ancillary data to provide a secondary, Eddington-unbiased, mass proxy, and shows that additional information, e.g. from optical surveys, can be exploited to keep the purity high at low SZ detection significance. 
In combination with on-going and future large optical and Near-IR surveys such as PanSTARRS, WISE, Euclid and LSST, and X-ray survey missions such as eROSITA, the final \textit{Planck} maps can thus be explored down to lower significance to provide a more complete accounting of haloes in the $z\gtrsim0.5$ regime.  

\begin{acknowledgements}
We thank Emanuele Daddi and Adam Muzzin for insightful discussions, and an anonymous referee for detailed suggestions that led to a substantial clarification of the paper. The research leading to these results has received funding from the European Research Council under the European Union's Seventh Framework Programme (FP7/2007-2013) / ERC grant agreement n$^{\circ}$ 340519. RB, AF and AS acknowledge financial support from the Spanish Ministry of Economy and Competitiveness (MINECO) under the 2011 Severo Ochoa Program MINECO SEV-2011-0187. HD acknowledges support from the Research Council of Norway. RH acknowledges support from the European Research Council FP7 grant number 279396. EP acknowledges the support of the French Agence Nationale de la Recherche under grant ANR-11-BS56-015.

Based on observations obtained with MegaPrime/MegaCam, a joint project of CFHT and CEA/DAPNIA, at the Canada-France-Hawaii Telescope (CFHT) which is operated by the National Research Council (NRC) of Canada, the Institute National des Sciences de l'Univers of the Centre National de la Recherche Scientifique of France, and the University of Hawaii.

This article is based on observations made with the Gran Telescopio Canarias (GTC), installed in the Spanish Observatorio del Roque de los Muchachos (ORM) of the Instituto de Astrof\'{\i}sica de Canarias (IAC), in the island of La Palma. This research used telescope time awarded by the CCI International Time Programme at the Canary Islands Observatories (ITP13-8).

The development of Planck has been supported by: ESA; CNES and CNRS/INSU-IN2P3-INP (France); ASI, CNR, and INAF (Italy); NASA and DoE (USA); STFC and UKSA (UK); CSIC, MICINN and JA (Spain); Tekes, AoF and CSC (Finland); DLR and MPG (Germany); CSA (Canada); DTU Space (Denmark); SER/SSO (Switzerland); RCN (Norway); SFI (Ireland); FCT/MCTES (Portugal); and The development of Planck has been supported by: ESA; CNES and CNRS/INSU-IN2P3-INP (France); ASI, CNR, and INAF (Italy); NASA and DoE (USA); STFC and UKSA (UK); CSIC, MICINN and JA (Spain); Tekes, AoF and CSC (Finland); DLR and MPG (Germany); CSA (Canada); DTU Space (Denmark); SER/SSO (Switzerland); RCN (Norway); SFI (Ireland); FCT/MCTES (Portugal); and PRACE (EU).

This research has made use of the following databases: the NED and IRSA databases, operated by the Jet Propulsion Laboratory, California Institute of Technology, under contract with the NASA; SIMBAD, operated at CDS, Strasbourg, France; SZ-cluster database operated by IDOC at IAS under contract with CNES and CNRS.
\end{acknowledgements}

\bibliographystyle{aa} 
\bibliography{MasterRefs} 

\begin{thebibliography}{62}
\expandafter\ifx\csname natexlab\endcsname\relax\def\natexlab#1{#1}\fi

\bibitem[{{Abell} {et~al.}(1989){Abell}, {Corwin}, \& {Olowin}}]{abell89}
{Abell}, G.~O., {Corwin}, Jr., H.~G., \& {Olowin}, R.~P. 1989, \apjs, 70, 1

\bibitem[{{Aghanim} {et~al.}(2015){Aghanim}, {Hurier}, {Diego}, {Douspis},
  {Macias-Perez}, {Pointecouteau}, {Comis}, {Arnaud}, \& {Montier}}]{aghanim15}
{Aghanim}, N., {Hurier}, G., {Diego}, J.-M., {et~al.} 2015, \aap, 580, A138

\bibitem[{{Ahn} {et~al.}(2012){Ahn}, {Alexandroff}, {Allende Prieto},
  {Anderson}, {Anderton}, {Andrews}, {Aubourg}, {Bailey}, {Balbinot}, {Barnes},
  \& et~al.}]{sdssdr9}
{Ahn}, C.~P., {Alexandroff}, R., {Allende Prieto}, C., {et~al.} 2012, \apjs,
  203, 21

\bibitem[{{Alam} {et~al.}(2015){Alam}, {Albareti}, {Allende Prieto}, {Anders},
  {Anderson}, {Anderton}, {Andrews}, {Armengaud}, {Aubourg}, {Bailey}, \&
  et~al.}]{sdssdr12}
{Alam}, S., {Albareti}, F.~D., {Allende Prieto}, C., {et~al.} 2015, \apjs, 219,
  12

\bibitem[{{Allen} {et~al.}(2011){Allen}, {Evrard}, \& {Mantz}}]{allen11}
{Allen}, S.~W., {Evrard}, A.~E., \& {Mantz}, A.~B. 2011, \araa, 49, 409

\bibitem[{{Andreon} \& {Congdon}(2014)}]{andreoncongdon}
{Andreon}, S. \& {Congdon}, P. 2014, \aap, 568, A23

\bibitem[{{Barkhouse} {et~al.}(2007){Barkhouse}, {Yee}, \&
  {L{\'o}pez-Cruz}}]{barkhouse07}
{Barkhouse}, W.~A., {Yee}, H.~K.~C., \& {L{\'o}pez-Cruz}, O. 2007, \apj, 671,
  1471

\bibitem[{{Benson} {et~al.}(2013){Benson}, {de Haan}, {Dudley}, {Reichardt},
  {Aird}, {Andersson}, {Armstrong}, {Ashby}, {Bautz}, {Bayliss}, {Bazin},
  {Bleem}, {Brodwin}, {Carlstrom}, {Chang}, {Cho}, {Clocchiatti}, {Crawford},
  {Crites}, {Desai}, {Dobbs}, {Foley}, {Forman}, {George}, {Gladders},
  {Gonzalez}, {Halverson}, {Harrington}, {High}, {Holder}, {Holzapfel},
  {Hoover}, {Hrubes}, {Jones}, {Joy}, {Keisler}, {Knox}, {Lee}, {Leitch},
  {Liu}, {Lueker}, {Luong-Van}, {Mantz}, {Marrone}, {McDonald}, {McMahon},
  {Mehl}, {Meyer}, {Mocanu}, {Mohr}, {Montroy}, {Murray}, {Natoli}, {Padin},
  {Plagge}, {Pryke}, {Rest}, {Ruel}, {Ruhl}, {Saliwanchik}, {Saro}, {Sayre},
  {Schaffer}, {Shaw}, {Shirokoff}, {Song}, {Spieler}, {Stalder},
  {Staniszewski}, {Stark}, {Story}, {Stubbs}, {Suhada}, {van Engelen},
  {Vanderlinde}, {Vieira}, {Vikhlinin}, {Williamson}, {Zahn}, \&
  {Zenteno}}]{benson13}
{Benson}, B.~A., {de Haan}, T., {Dudley}, J.~P., {et~al.} 2013, \apj, 763, 147

\bibitem[{{Bertin}(2006)}]{scamp}
{Bertin}, E. 2006, in Astronomical Society of the Pacific Conference Series,
  Vol. 351, Astronomical Data Analysis Software and Systems XV, ed.
  C.~{Gabriel}, C.~{Arviset}, D.~{Ponz}, \& S.~{Enrique}, 112

\bibitem[{{Bertin}(2011)}]{psfex}
{Bertin}, E. 2011, in Astronomical Society of the Pacific Conference Series,
  Vol. 442, Astronomical Data Analysis Software and Systems XX, ed. I.~N.
  {Evans}, A.~{Accomazzi}, D.~J. {Mink}, \& A.~H. {Rots}, 435

\bibitem[{{Bleem} {et~al.}(2015){Bleem}, {Stalder}, {de Haan}, {Aird}, {Allen},
  {Applegate}, {Ashby}, {Bautz}, {Bayliss}, {Benson}, {Bocquet}, {Brodwin},
  {Carlstrom}, {Chang}, {Chiu}, {Cho}, {Clocchiatti}, {Crawford}, {Crites},
  {Desai}, {Dietrich}, {Dobbs}, {Foley}, {Forman}, {George}, {Gladders},
  {Gonzalez}, {Halverson}, {Hennig}, {Hoekstra}, {Holder}, {Holzapfel},
  {Hrubes}, {Jones}, {Keisler}, {Knox}, {Lee}, {Leitch}, {Liu}, {Lueker},
  {Luong-Van}, {Mantz}, {Marrone}, {McDonald}, {McMahon}, {Meyer}, {Mocanu},
  {Mohr}, {Murray}, {Padin}, {Pryke}, {Reichardt}, {Rest}, {Ruel}, {Ruhl},
  {Saliwanchik}, {Saro}, {Sayre}, {Schaffer}, {Schrabback}, {Shirokoff},
  {Song}, {Spieler}, {Stanford}, {Staniszewski}, {Stark}, {Story}, {Stubbs},
  {Vanderlinde}, {Vieira}, {Vikhlinin}, {Williamson}, {Zahn}, \&
  {Zenteno}}]{bleem15}
{Bleem}, L.~E., {Stalder}, B., {de Haan}, T., {et~al.} 2015, \apjs, 216, 27

\bibitem[{{Bocquet} {et~al.}(2015){Bocquet}, {Saro}, {Dolag}, \&
  {Mohr}}]{bocquet15}
{Bocquet}, S., {Saro}, A., {Dolag}, K., \& {Mohr}, J.~J. 2015, ArXiv e-prints

\bibitem[{{B{\"o}hringer} {et~al.}(2004){B{\"o}hringer}, {Schuecker}, {Guzzo},
  {Collins}, {Voges}, {Cruddace}, {Ortiz-Gil}, {Chincarini}, {De Grandi},
  {Edge}, {MacGillivray}, {Neumann}, {Schindler}, \& {Shaver}}]{bohringer04}
{B{\"o}hringer}, H., {Schuecker}, P., {Guzzo}, L., {et~al.} 2004, \aap, 425,
  367

\bibitem[{{B{\"o}hringer} {et~al.}(2000){B{\"o}hringer}, {Voges}, {Huchra},
  {McLean}, {Giacconi}, {Rosati}, {Burg}, {Mader}, {Schuecker}, {Simi{\c c}},
  {Komossa}, {Reiprich}, {Retzlaff}, \& {Tr{\"u}mper}}]{bohringer00}
{B{\"o}hringer}, H., {Voges}, W., {Huchra}, J.~P., {et~al.} 2000, \apjs, 129,
  435

\bibitem[{{da Silva} {et~al.}(2004){da Silva}, {Kay}, {Liddle}, \&
  {Thomas}}]{dasilva04}
{da Silva}, A.~C., {Kay}, S.~T., {Liddle}, A.~R., \& {Thomas}, P.~A. 2004,
  \mnras, 348, 1401

\bibitem[{{Ebeling} {et~al.}(2007){Ebeling}, {Barrett}, {Donovan}, {Ma},
  {Edge}, \& {van Speybroeck}}]{ebeling07}
{Ebeling}, H., {Barrett}, E., {Donovan}, D., {et~al.} 2007, \apjl, 661, L33

\bibitem[{{Ebeling} {et~al.}(1998){Ebeling}, {Edge}, {Bohringer}, {Allen},
  {Crawford}, {Fabian}, {Voges}, \& {Huchra}}]{ebeling98}
{Ebeling}, H., {Edge}, A.~C., {Bohringer}, H., {et~al.} 1998, \mnras, 301, 881

\bibitem[{{Eddington}(1913)}]{eddington1913}
{Eddington}, A.~S. 1913, \mnras, 73, 359

\bibitem[{{Erben} {et~al.}(2009){Erben}, {Hildebrandt}, {Lerchster}, {Hudelot},
  {Benjamin}, {van Waerbeke}, {Schrabback}, {Brimioulle}, {Cordes}, {Dietrich},
  {Holhjem}, {Schirmer}, \& {Schneider}}]{erben09}
{Erben}, T., {Hildebrandt}, H., {Lerchster}, M., {et~al.} 2009, \aap, 493, 1197

\bibitem[{{Fassbender} {et~al.}(2011){Fassbender}, {B{\"o}hringer}, {Nastasi},
  {{\v S}uhada}, {M{\"u}hlegger}, {de Hoon}, {Kohnert}, {Lamer}, {Mohr},
  {Pierini}, {Pratt}, {Quintana}, {Rosati}, {Santos}, \&
  {Schwope}}]{fassbender11}
{Fassbender}, R., {B{\"o}hringer}, H., {Nastasi}, A., {et~al.} 2011, New
  Journal of Physics, 13, 125014

\bibitem[{{Gettings} {et~al.}(2012){Gettings}, {Gonzalez}, {Stanford},
  {Eisenhardt}, {Brodwin}, {Mancone}, {Stern}, {Zeimann}, {Masci}, {Papovich},
  {Tanaka}, \& {Wright}}]{gettings12}
{Gettings}, D.~P., {Gonzalez}, A.~H., {Stanford}, S.~A., {et~al.} 2012, \apjl,
  759, L23

\bibitem[{{Gilbank} {et~al.}(2011){Gilbank}, {Gladders}, {Yee}, \&
  {Hsieh}}]{gilbank11}
{Gilbank}, D.~G., {Gladders}, M.~D., {Yee}, H.~K.~C., \& {Hsieh}, B.~C. 2011,
  \aj, 141, 94

\bibitem[{{Gladders} \& {Yee}(2000)}]{gladdersyee00}
{Gladders}, M.~D. \& {Yee}, H.~K.~C. 2000, \aj, 120, 2148

\bibitem[{{Hasselfield} {et~al.}(2013){Hasselfield}, {Hilton}, {Marriage},
  {Addison}, {Barrientos}, {Battaglia}, {Battistelli}, {Bond}, {Crichton},
  {Das}, {Devlin}, {Dicker}, {Dunkley}, {D{\"u}nner}, {Fowler}, {Gralla},
  {Hajian}, {Halpern}, {Hincks}, {Hlozek}, {Hughes}, {Infante}, {Irwin},
  {Kosowsky}, {Marsden}, {Menanteau}, {Moodley}, {Niemack}, {Nolta}, {Page},
  {Partridge}, {Reese}, {Schmitt}, {Sehgal}, {Sherwin}, {Sievers}, {Sif{\'o}n},
  {Spergel}, {Staggs}, {Swetz}, {Switzer}, {Thornton}, {Trac}, \&
  {Wollack}}]{hasselfield13}
{Hasselfield}, M., {Hilton}, M., {Marriage}, T.~A., {et~al.} 2013, \jcap, 7, 8

\bibitem[{{Hildebrandt} {et~al.}(2009){Hildebrandt}, {Pielorz}, {Erben}, {van
  Waerbeke}, {Simon}, \& {Capak}}]{hildebrandt09a}
{Hildebrandt}, H., {Pielorz}, J., {Erben}, T., {et~al.} 2009, \aap, 498, 725

\bibitem[{{Hoekstra} {et~al.}(2015){Hoekstra}, {Herbonnet}, {Muzzin}, {Babul},
  {Mahdavi}, {Viola}, \& {Cacciato}}]{hoekstra15}
{Hoekstra}, H., {Herbonnet}, R., {Muzzin}, A., {et~al.} 2015, \mnras, 449, 685

\bibitem[{{Hoekstra} {et~al.}(2012){Hoekstra}, {Mahdavi}, {Babul}, \&
  {Bildfell}}]{hoekstra12}
{Hoekstra}, H., {Mahdavi}, A., {Babul}, A., \& {Bildfell}, C. 2012, \mnras,
  427, 1298

\bibitem[{{Ilbert} {et~al.}(2013){Ilbert}, {McCracken}, {Le F{\`e}vre},
  {Capak}, {Dunlop}, {Karim}, {Renzini}, {Caputi}, {Boissier}, {Arnouts},
  {Aussel}, {Comparat}, {Guo}, {Hudelot}, {Kartaltepe}, {Kneib}, {Krogager},
  {Le Floc'h}, {Lilly}, {Mellier}, {Milvang-Jensen}, {Moutard}, {Onodera},
  {Richard}, {Salvato}, {Sanders}, {Scoville}, {Silverman}, {Taniguchi},
  {Tasca}, {Thomas}, {Toft}, {Tresse}, {Vergani}, {Wolk}, \& {Zirm}}]{ilbert13}
{Ilbert}, O., {McCracken}, H.~J., {Le F{\`e}vre}, O., {et~al.} 2013, \aap, 556,
  A55

\bibitem[{{Kirk} {et~al.}(2015){Kirk}, {Hilton}, {Cress}, {Crawford}, {Hughes},
  {Battaglia}, {Bond}, {Burke}, {Gralla}, {Hajian}, {Hasselfield}, {Hincks},
  {Infante}, {Kosowsky}, {Marriage}, {Menanteau}, {Moodley}, {Niemack},
  {Sievers}, {Sif{\'o}n}, {Wilson}, {Wollack}, \& {Zunckel}}]{kirk15}
{Kirk}, B., {Hilton}, M., {Cress}, C., {et~al.} 2015, \mnras, 449, 4010

\bibitem[{{Liu} {et~al.}(2015){Liu}, {Hennig}, {Desai}, {Hoyle},
  {Koppenhoefer}, {Mohr}, {Paech}, {Burgett}, {Chambers}, {Cole}, {Draper},
  {Kaiser}, {Metcalfe}, {Morgan}, {Price}, {Stubbs}, {Tonry}, {Wainscoat}, \&
  {Waters}}]{liu15}
{Liu}, J., {Hennig}, C., {Desai}, S., {et~al.} 2015, \mnras, 449, 3370

\bibitem[{{Magnier} \& {Cuillandre}(2004)}]{elixir04}
{Magnier}, E.~A. \& {Cuillandre}, J.-C. 2004, \pasp, 116, 449

\bibitem[{{Mantz} {et~al.}(2010{\natexlab{a}}){Mantz}, {Allen}, {Ebeling},
  {Rapetti}, \& {Drlica-Wagner}}]{mantz10}
{Mantz}, A., {Allen}, S.~W., {Ebeling}, H., {Rapetti}, D., \& {Drlica-Wagner},
  A. 2010{\natexlab{a}}, \mnras, 406, 1773

\bibitem[{{Mantz} {et~al.}(2010{\natexlab{b}}){Mantz}, {Allen}, {Rapetti}, \&
  {Ebeling}}]{mantz10a}
{Mantz}, A., {Allen}, S.~W., {Rapetti}, D., \& {Ebeling}, H.
  2010{\natexlab{b}}, \mnras, 406, 1759

\bibitem[{{Melin} {et~al.}(2006){Melin}, {Bartlett}, \&
  {Delabrouille}}]{melin06}
{Melin}, J.-B., {Bartlett}, J.~G., \& {Delabrouille}, J. 2006, \aap, 459, 341

\bibitem[{{Moretti} {et~al.}(2015){Moretti}, {Bettoni}, {Poggianti}, {Fasano},
  {Varela}, {D'Onofrio}, {Vulcani}, {Cava}, {Fritz}, {Couch}, {Moles}, \&
  {Kj{\ae}rgaard}}]{moretti15}
{Moretti}, A., {Bettoni}, D., {Poggianti}, B.~M., {et~al.} 2015, \aap, 581, A11

\bibitem[{{Muzzin} {et~al.}(2013{\natexlab{a}}){Muzzin}, {Marchesini},
  {Stefanon}, {Franx}, {McCracken}, {Milvang-Jensen}, {Dunlop}, {Fynbo},
  {Brammer}, {Labb{\'e}}, \& {van Dokkum}}]{muzzin13b}
{Muzzin}, A., {Marchesini}, D., {Stefanon}, M., {et~al.} 2013{\natexlab{a}},
  \apj, 777, 18

\bibitem[{{Muzzin} {et~al.}(2013{\natexlab{b}}){Muzzin}, {Marchesini},
  {Stefanon}, {Franx}, {Milvang-Jensen}, {Dunlop}, {Fynbo}, {Brammer},
  {Labb{\'e}}, \& {van Dokkum}}]{muzzin13a}
{Muzzin}, A., {Marchesini}, D., {Stefanon}, M., {et~al.} 2013{\natexlab{b}},
  \apjs, 206, 8

\bibitem[{{Planck Collaboration} {et~al.}(2014{\natexlab{a}}){Planck
  Collaboration}, {Ade}, {Aghanim}, {Armitage-Caplan}, {Arnaud}, {Ashdown},
  {Atrio-Barandela}, {Aumont}, {Aussel}, {Baccigalupi}, \& et~al.}]{psz1}
{Planck Collaboration}, {Ade}, P.~A.~R., {Aghanim}, N., {et~al.}
  2014{\natexlab{a}}, \aap, 571, A29

\bibitem[{{Planck Collaboration} {et~al.}(2015{\natexlab{a}}){Planck
  Collaboration}, {Ade}, {Aghanim}, {Armitage-Caplan}, {Arnaud}, {Ashdown},
  {Atrio-Barandela}, {Aumont}, {Aussel}, {Baccigalupi}, \&
  et~al.}]{planckpsz1v215}
{Planck Collaboration}, {Ade}, P.~A.~R., {Aghanim}, N., {et~al.}
  2015{\natexlab{a}}, ArXiv e-prints [1502.00543]

\bibitem[{{Planck Collaboration} {et~al.}(2014{\natexlab{b}}){Planck
  Collaboration}, {Ade}, {Aghanim}, {Armitage-Caplan}, {Arnaud}, {Ashdown},
  {Atrio-Barandela}, {Aumont}, {Baccigalupi}, {Banday}, \&
  et~al.}]{planck13cosmology}
{Planck Collaboration}, {Ade}, P.~A.~R., {Aghanim}, N., {et~al.}
  2014{\natexlab{b}}, \aap, 571, A16

\bibitem[{{Planck Collaboration} {et~al.}(2014{\natexlab{c}}){Planck
  Collaboration}, {Ade}, {Aghanim}, {Armitage-Caplan}, {Arnaud}, {Ashdown},
  {Atrio-Barandela}, {Aumont}, {Baccigalupi}, {Banday}, \&
  et~al.}]{planckclustercosmo14}
{Planck Collaboration}, {Ade}, P.~A.~R., {Aghanim}, N., {et~al.}
  2014{\natexlab{c}}, \aap, 571, A20

\bibitem[{{Planck Collaboration} {et~al.}(2011{\natexlab{a}}){Planck
  Collaboration}, {Ade}, {Aghanim}, {Arnaud}, {Ashdown}, {Aumont},
  {Baccigalupi}, {Balbi}, {Banday}, {Barreiro}, \& et~al.}]{PlanckESZ}
{Planck Collaboration}, {Ade}, P.~A.~R., {Aghanim}, N., {et~al.}
  2011{\natexlab{a}}, \aap, 536, A8

\bibitem[{{Planck Collaboration} {et~al.}(2015{\natexlab{b}}){Planck
  Collaboration}, {Ade}, {Aghanim}, {Arnaud}, {Ashdown}, {Aumont},
  {Baccigalupi}, {Banday}, {Barreiro}, {Barrena}, {Bartolo}, {Battaner},
  {Benabed}, {Benoit-L{\'e}vy}, {Bernard}, {Bersanelli}, {Bielewicz},
  {Bikmaev}, {B{\"o}hringer}, {Bonaldi}, {Bonavera}, {Bond}, {Borrill},
  {Bouchet}, {Burenin}, {Burigana}, {Butler}, {Calabrese}, {Carvalho},
  {Catalano}, {Chamballu}, {Chiang}, {Chon}, {Christensen}, {Churazov},
  {Clements}, {Colombo}, {Comis}, {Couchot}, {Curto}, {Cuttaia}, {Dahle},
  {Danese}, {Davies}, {Davis}, {de Bernardis}, {de Rosa}, {de Zotti},
  {Delabrouille}, {Diego}, {Dole}, {Dor{\'e}}, {Douspis}, {Ducout}, {Dupac},
  {Efstathiou}, {Elsner}, {En{\ss}lin}, {Eriksen}, {Finelli}, {Flores-Cacho},
  {Forni}, {Frailis}, {Fraisse}, {Franceschi}, {Frejsel}, {Fromenteau},
  {Galeotta}, {Ganga}, {G{\'e}nova-Santos}, {Giard}, {Gilfanov},
  {Giraud-H{\'e}raud}, {Gjerl{\o}w}, {Gonz{\'a}lez-Nuevo}, {G{\'o}rski},
  {Gruppuso}, {Hansen}, {Hanson}, {Harrison}, {Hempel}, {Henrot-Versill{\'e}},
  {Hern{\'a}ndez-Monteagudo}, {Herranz}, {Hildebrandt}, {Hivon}, {Hobson},
  {Holmes}, {Hornstrup}, {Hovest}, {Huffenberger}, {Hurier}, {Jaffe}, {Jones},
  {Juvela}, {Keih{\"a}nen}, {Keskitalo}, {Khamitov}, {Kisner}, {Kneissl},
  {Knoche}, {Kunz}, {Kurki-Suonio}, {Lagache}, {Lamarre}, {Lasenby},
  {Lattanzi}, {Lawrence}, {Leonardi}, {Levrier}, {Liguori}, {Lilje},
  {Linden-V{\o}rnle}, {L{\'o}pez-Caniego}, {Lubin}, {Mac{\'{\i}}as-P{\'e}rez},
  {Maino}, {Mandolesi}, {Maris}, {Martin}, {Mart{\'{\i}}nez-Gonz{\'a}lez},
  {Masi}, {Matarrese}, {Mazzotta}, {Melin}, {Mendes}, {Mennella}, {Migliaccio},
  {Miville-Desch{\^e}nes}, {Moneti}, {Montier}, {Morgante}, {Mortlock},
  {Munshi}, {Murphy}, {Naselsky}, {Nati}, {Natoli}, {N{\o}rgaard-Nielsen},
  {Novikov}, {Novikov}, {Oxborrow}, {Pagano}, {Pajot}, {Paoletti}, {Pasian},
  {Perdereau}, {Perotto}, {Perrotta}, {Pettorino}, {Piacentini}, {Piat},
  {Pietrobon}, {Plaszczynski}, {Pointecouteau}, {Polenta}, {Popa}, {Pratt},
  {Prunet}, {Puget}, {Rachen}, {Reinecke}, {Remazeilles}, {Renault},
  {Ricciardi}, {Ristorcelli}, {Rocha}, {Roman}, {Rosset}, {Rossetti},
  {Roudier}, {Rubi{\~n}o-Mart{\'{\i}}n}, {Rusholme}, {Sandri}, {Scott},
  {Spencer}, {Stolyarov}, {Sudiwala}, {Sunyaev}, {Sutton}, {Suur-Uski},
  {Sygnet}, {Tauber}, {Terenzi}, {Toffolatti}, {Tomasi}, {Tristram}, {Tucci},
  {Valenziano}, {Valiviita}, {Van Tent}, {Vibert}, {Vielva}, {Villa}, {Wade},
  {Wandelt}, {Wehus}, {Yvon}, {Zacchei}, \& {Zonca}}]{planckrtt15}
{Planck Collaboration}, {Ade}, P.~A.~R., {Aghanim}, N., {et~al.}
  2015{\natexlab{b}}, \aap, 582, A29

\bibitem[{{Planck Collaboration} {et~al.}(2015{\natexlab{c}}){Planck
  Collaboration}, {Ade}, {Aghanim}, {Arnaud}, {Ashdown}, {Aumont},
  {Baccigalupi}, {Banday}, {Barreiro}, {Barrena}, {Bartolo}, {Battaner},
  {Benabed}, {Benoit-L{\'e}vy}, {Bernard}, {Bersanelli}, {Bielewicz},
  {Bikmaev}, {B{\"o}hringer}, {Bonaldi}, {Bonavera}, {Bond}, {Borrill},
  {Bouchet}, {Burenin}, {Burigana}, {Calabrese}, {Cardoso}, {Catalano},
  {Chamballu}, {Chary}, {Chiang}, {Chon}, {Christensen}, {Clements}, {Colombo},
  {Combet}, {Comis}, {Crill}, {Curto}, {Cuttaia}, {Dahle}, {Danese}, {Davies},
  {Davis}, {de Bernardis}, {de Rosa}, {de Zotti}, {Delabrouille}, {Diego},
  {Dole}, {Donzelli}, {Dor{\'e}}, {Douspis}, {Dupac}, {Efstathiou}, {Elsner},
  {En{\ss}lin}, {Eriksen}, {Ferragamo}, {Finelli}, {Forni}, {Frailis},
  {Fraisse}, {Franceschi}, {Fromenteau}, {Galeotta}, {Galli}, {Ganga},
  {G{\'e}nova-Santos}, {Giard}, {Gjerl{\o}w}, {Gonz{\'a}lez-Nuevo},
  {G{\'o}rski}, {Gruppuso}, {Hansen}, {Harrison}, {Hempel},
  {Hern{\'a}ndez-Monteagudo}, {Herranz}, {Hildebrandt}, {Hivon}, {Hornstrup},
  {Hovest}, {Huffenberger}, {Hurier}, {Jaffe}, {Keih{\"a}nen}, {Keskitalo},
  {Khamitov}, {Kisner}, {Knoche}, {Kunz}, {Kurki-Suonio}, {Lamarre}, {Lasenby},
  {Lattanzi}, {Lawrence}, {Leonardi}, {Le{\'o}n-Tavares}, {Levrier}, {Lietzen},
  {Liguori}, {Lilje}, {Linden-V{\o}rnle}, {L{\'o}pez-Caniego}, {Lubin},
  {Mac{\'{\i}}as-P{\'e}rez}, {Maffei}, {Maino}, {Mandolesi}, {Maris}, {Martin},
  {Mart{\'{\i}}nez-Gonz{\'a}lez}, {Masi}, {Matarrese}, {McGehee}, {Melchiorri},
  {Mennella}, {Migliaccio}, {Miville-Desch{\^e}nes}, {Moneti}, {Montier},
  {Morgante}, {Mortlock}, {Munshi}, {Murphy}, {Naselsky}, {Nati}, {Natoli},
  {Novikov}, {Novikov}, {Oxborrow}, {Pagano}, {Pajot}, {Paoletti}, {Pasian},
  {Perdereau}, {Pettorino}, {Piacentini}, {Piat}, {Pierpaoli}, {Plaszczynski},
  {Pointecouteau}, {Polenta}, {Pratt}, {Prunet}, {Puget}, {Rachen}, {Rebolo},
  {Reinecke}, {Remazeilles}, {Renault}, {Renzi}, {Ristorcelli}, {Rocha},
  {Rosset}, {Rossetti}, {Roudier}, {Rubi{\~n}o-Mart{\'{\i}}n}, {Rusholme},
  {Sandri}, {Santos}, {Savelainen}, {Savini}, {Scott}, {Stolyarov},
  {Streblyanska}, {Sudiwala}, {Sunyaev}, {Suur-Uski}, {Sygnet}, {Terenzi},
  {Toffolatti}, {Tomasi}, {Tramonte}, {Tristram}, {Tucci}, {Valenziano},
  {Valiviita}, {Van Tent}, {Vielva}, {Villa}, {Wade}, {Wandelt}, {Wehus},
  {Yvon}, {Zacchei}, \& {Zonca}}]{planckcanary15}
{Planck Collaboration}, {Ade}, P.~A.~R., {Aghanim}, N., {et~al.}
  2015{\natexlab{c}}, ArXiv e-prints [1504.04583]

\bibitem[{{Planck Collaboration} {et~al.}(2015{\natexlab{d}}){Planck
  Collaboration}, {Ade}, {Aghanim}, {Arnaud}, {Ashdown}, {Aumont},
  {Baccigalupi}, {Banday}, {Barreiro}, {Barrena}, \& et~al.}]{psz2}
{Planck Collaboration}, {Ade}, P.~A.~R., {Aghanim}, N., {et~al.}
  2015{\natexlab{d}}, ArXiv e-prints [1502.01598]

\bibitem[{{Planck Collaboration} {et~al.}(2015{\natexlab{e}}){Planck
  Collaboration}, {Ade}, {Aghanim}, {Arnaud}, {Ashdown}, {Aumont},
  {Baccigalupi}, {Banday}, {Barreiro}, {Bartlett}, \&
  et~al.}]{planck15clustercosmology}
{Planck Collaboration}, {Ade}, P.~A.~R., {Aghanim}, N., {et~al.}
  2015{\natexlab{e}}, ArXiv e-prints [1502.01597]

\bibitem[{{Planck Collaboration} {et~al.}(2011{\natexlab{b}}){Planck
  Collaboration}, {Aghanim}, {Arnaud}, {Ashdown}, {Aumont}, {Baccigalupi},
  {Balbi}, {Banday}, {Barreiro}, {Bartelmann}, \& et~al.}]{planckxmm11}
{Planck Collaboration}, {Aghanim}, N., {Arnaud}, M., {et~al.}
  2011{\natexlab{b}}, \aap, 536, A9

\bibitem[{{Pratt} {et~al.}(2009){Pratt}, {Croston}, {Arnaud}, \&
  {B{\"o}hringer}}]{pratt09}
{Pratt}, G.~W., {Croston}, J.~H., {Arnaud}, M., \& {B{\"o}hringer}, H. 2009,
  \aap, 498, 361

\bibitem[{{Reichardt} {et~al.}(2013){Reichardt}, {Stalder}, {Bleem}, {Montroy},
  {Aird}, {Andersson}, {Armstrong}, {Ashby}, {Bautz}, {Bayliss}, {Bazin},
  {Benson}, {Brodwin}, {Carlstrom}, {Chang}, {Cho}, {Clocchiatti}, {Crawford},
  {Crites}, {de Haan}, {Desai}, {Dobbs}, {Dudley}, {Foley}, {Forman}, {George},
  {Gladders}, {Gonzalez}, {Halverson}, {Harrington}, {High}, {Holder},
  {Holzapfel}, {Hoover}, {Hrubes}, {Jones}, {Joy}, {Keisler}, {Knox}, {Lee},
  {Leitch}, {Liu}, {Lueker}, {Luong-Van}, {Mantz}, {Marrone}, {McDonald},
  {McMahon}, {Mehl}, {Meyer}, {Mocanu}, {Mohr}, {Murray}, {Natoli}, {Padin},
  {Plagge}, {Pryke}, {Rest}, {Ruel}, {Ruhl}, {Saliwanchik}, {Saro}, {Sayre},
  {Schaffer}, {Shaw}, {Shirokoff}, {Song}, {Spieler}, {Staniszewski}, {Stark},
  {Story}, {Stubbs}, {{\v S}uhada}, {van Engelen}, {Vanderlinde}, {Vieira},
  {Vikhlinin}, {Williamson}, {Zahn}, \& {Zenteno}}]{reichardt13}
{Reichardt}, C.~L., {Stalder}, B., {Bleem}, L.~E., {et~al.} 2013, \apj, 763,
  127

\bibitem[{{Rozo} {et~al.}(2015){Rozo}, {Rykoff}, {Bartlett}, \&
  {Melin}}]{rozo15}
{Rozo}, E., {Rykoff}, E.~S., {Bartlett}, J.~G., \& {Melin}, J.-B. 2015, \mnras,
  450, 592

\bibitem[{{Rozo} {et~al.}(2010){Rozo}, {Wechsler}, {Rykoff}, {Annis}, {Becker},
  {Evrard}, {Frieman}, {Hansen}, {Hao}, {Johnston}, {Koester}, {McKay},
  {Sheldon}, \& {Weinberg}}]{rozo10}
{Rozo}, E., {Wechsler}, R.~H., {Rykoff}, E.~S., {et~al.} 2010, \apj, 708, 645

\bibitem[{{Rykoff} {et~al.}(2014){Rykoff}, {Rozo}, {Busha}, {Cunha},
  {Finoguenov}, {Evrard}, {Hao}, {Koester}, {Leauthaud}, {Nord}, {Pierre},
  {Reddick}, {Sadibekova}, {Sheldon}, \& {Wechsler}}]{rykoff14}
{Rykoff}, E.~S., {Rozo}, E., {Busha}, M.~T., {et~al.} 2014, \apj, 785, 104

\bibitem[{{Schlegel} {et~al.}(1998){Schlegel}, {Finkbeiner}, \&
  {Davis}}]{schlegel98}
{Schlegel}, D.~J., {Finkbeiner}, D.~P., \& {Davis}, M. 1998, \apj, 500, 525

\bibitem[{{Sehgal} {et~al.}(2011){Sehgal}, {Trac}, {Acquaviva}, {Ade},
  {Aguirre}, {Amiri}, {Appel}, {Barrientos}, {Battistelli}, {Bond}, {Brown},
  {Burger}, {Chervenak}, {Das}, {Devlin}, {Dicker}, {Bertrand Doriese},
  {Dunkley}, {D{\"u}nner}, {Essinger-Hileman}, {Fisher}, {Fowler}, {Hajian},
  {Halpern}, {Hasselfield}, {Hern{\'a}ndez-Monteagudo}, {Hilton}, {Hilton},
  {Hincks}, {Hlozek}, {Holtz}, {Huffenberger}, {Hughes}, {Hughes}, {Infante},
  {Irwin}, {Jones}, {Baptiste Juin}, {Klein}, {Kosowsky}, {Lau}, {Limon},
  {Lin}, {Lupton}, {Marriage}, {Marsden}, {Martocci}, {Mauskopf}, {Menanteau},
  {Moodley}, {Moseley}, {Netterfield}, {Niemack}, {Nolta}, {Page}, {Parker},
  {Partridge}, {Reid}, {Sherwin}, {Sievers}, {Spergel}, {Staggs}, {Swetz},
  {Switzer}, {Thornton}, {Tucker}, {Warne}, {Wollack}, \& {Zhao}}]{sehgal11}
{Sehgal}, N., {Trac}, H., {Acquaviva}, V., {et~al.} 2011, \apj, 732, 44

\bibitem[{{Sunyaev} \& {Zeldovich}(1980)}]{sunyaev80}
{Sunyaev}, R.~A. \& {Zeldovich}, I.~B. 1980, \araa, 18, 537

\bibitem[{{Tinker} {et~al.}(2008){Tinker}, {Kravtsov}, {Klypin}, {Abazajian},
  {Warren}, {Yepes}, {Gottl{\"o}ber}, \& {Holz}}]{tinker08}
{Tinker}, J., {Kravtsov}, A.~V., {Klypin}, A., {et~al.} 2008, \apj, 688, 709

\bibitem[{{van der Burg} {et~al.}(2015){van der Burg}, {Hoekstra}, {Muzzin},
  {Sif{\'o}n}, {Balogh}, \& {McGee}}]{vdB15}
{van der Burg}, R.~F.~J., {Hoekstra}, H., {Muzzin}, A., {et~al.} 2015, \aap,
  577, A19

\bibitem[{{van der Burg} {et~al.}(2013){van der Burg}, {Muzzin}, {Hoekstra},
  {Lidman}, {Rettura}, {Wilson}, {Yee}, {Hildebrandt}, {Marchesini},
  {Stefanon}, {Demarco}, \& {Kuijken}}]{vdB13}
{van der Burg}, R.~F.~J., {Muzzin}, A., {Hoekstra}, H., {et~al.} 2013, \aap,
  557, A15

\bibitem[{{van Dokkum}(2001)}]{dokkumcosmics}
{van Dokkum}, P.~G. 2001, \pasp, 113, 1420

\bibitem[{{Vikhlinin} {et~al.}(2009){Vikhlinin}, {Kravtsov}, {Burenin},
  {Ebeling}, {Forman}, {Hornstrup}, {Jones}, {Murray}, {Nagai}, {Quintana}, \&
  {Voevodkin}}]{vikhlinin09}
{Vikhlinin}, A., {Kravtsov}, A.~V., {Burenin}, R.~A., {et~al.} 2009, \apj, 692,
  1060

\bibitem[{{Williams} {et~al.}(2009){Williams}, {Quadri}, {Franx}, {van Dokkum},
  \& {Labb{\'e}}}]{williams09}
{Williams}, R.~J., {Quadri}, R.~F., {Franx}, M., {van Dokkum}, P., \&
  {Labb{\'e}}, I. 2009, \apj, 691, 1879

\bibitem[{{Wright} {et~al.}(2010){Wright}, {Eisenhardt}, {Mainzer}, {Ressler},
  {Cutri}, {Jarrett}, {Kirkpatrick}, {Padgett}, {McMillan}, {Skrutskie},
  {Stanford}, {Cohen}, {Walker}, {Mather}, {Leisawitz}, {Gautier}, {McLean},
  {Benford}, {Lonsdale}, {Blain}, {Mendez}, {Irace}, {Duval}, {Liu}, {Royer},
  {Heinrichsen}, {Howard}, {Shannon}, {Kendall}, {Walsh}, {Larsen}, {Cardon},
  {Schick}, {Schwalm}, {Abid}, {Fabinsky}, {Naes}, \& {Tsai}}]{wise}
{Wright}, E.~L., {Eisenhardt}, P.~R.~M., {Mainzer}, A.~K., {et~al.} 2010, \aj,
  140, 1868

\end{thebibliography}

\begin{appendix}
\section{$r-z$-colour images for the targeted systems with an optical counterpart}\label{sec:images}

\begin{figure*}
\centering
\subfigure[\texttt{PLCK G027.65-34.27}]{\label{fig:mmf3-2689}\includegraphics[width=0.4\textwidth]{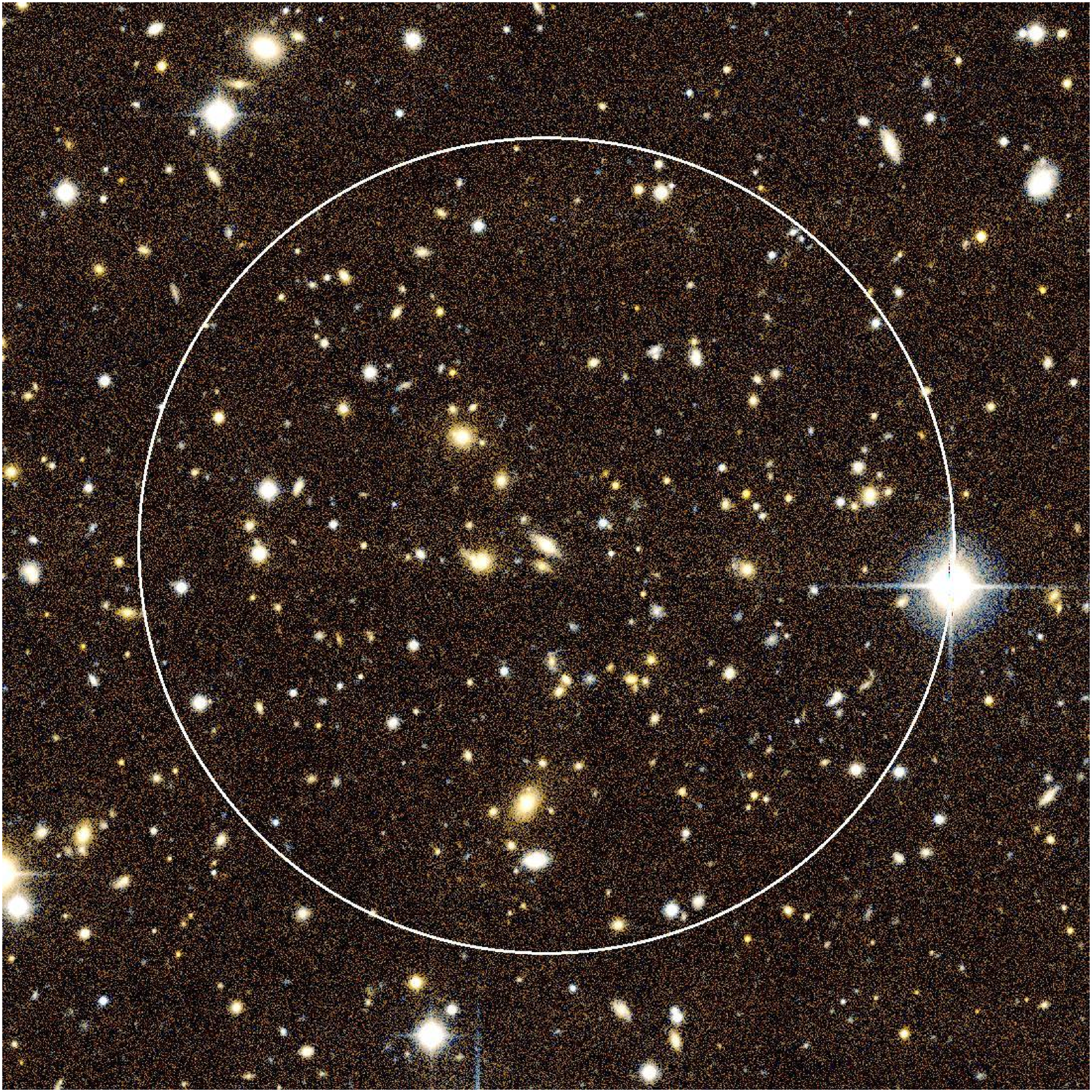}}
\subfigure[\texttt{PLCK G038.64-41.15}]{\label{fig:mmf3-2695}\includegraphics[width=0.4\textwidth]{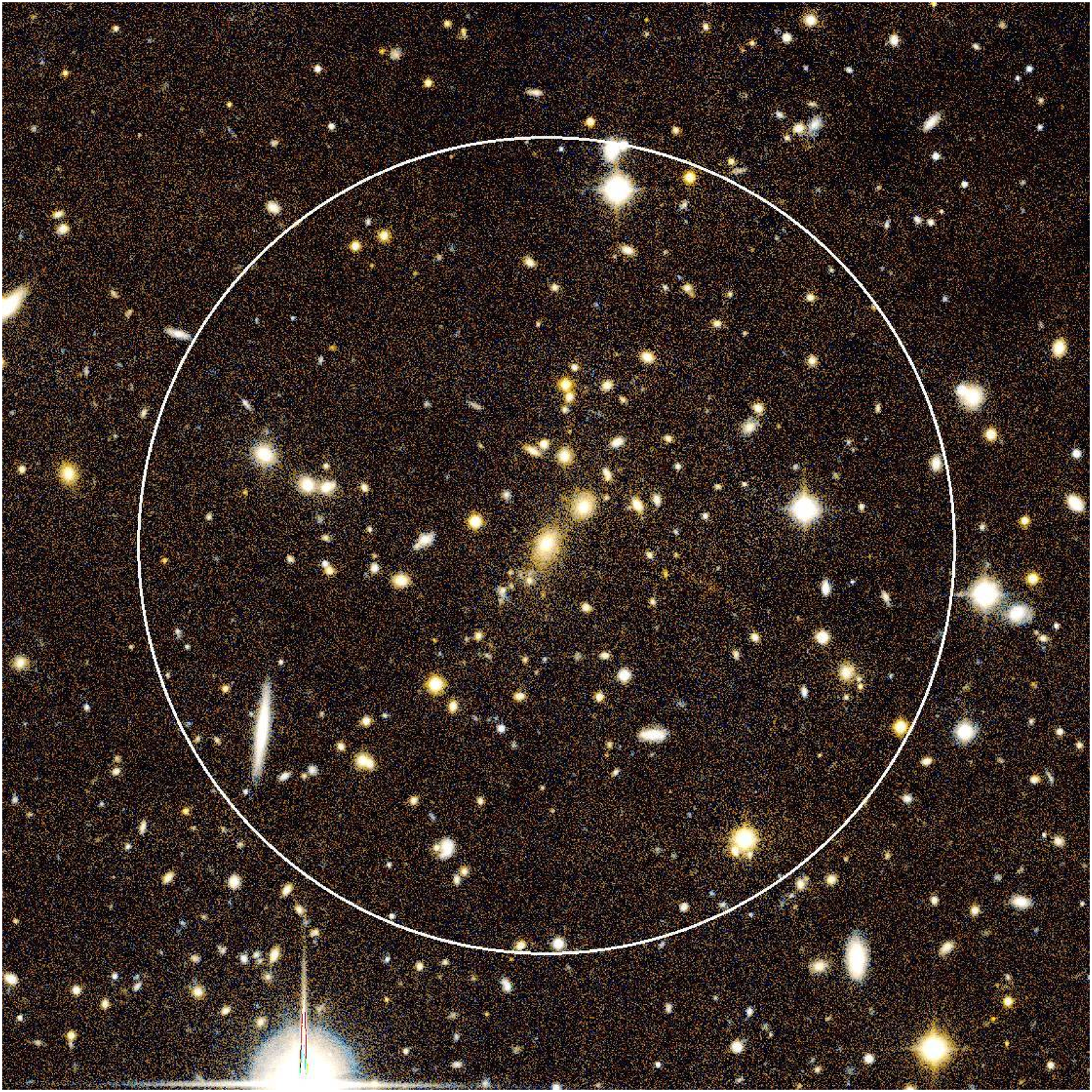}}
\subfigure[\texttt{PSZ2 G041.69+21.68}]{\label{fig:plck-1448}\includegraphics[width=0.4\textwidth]{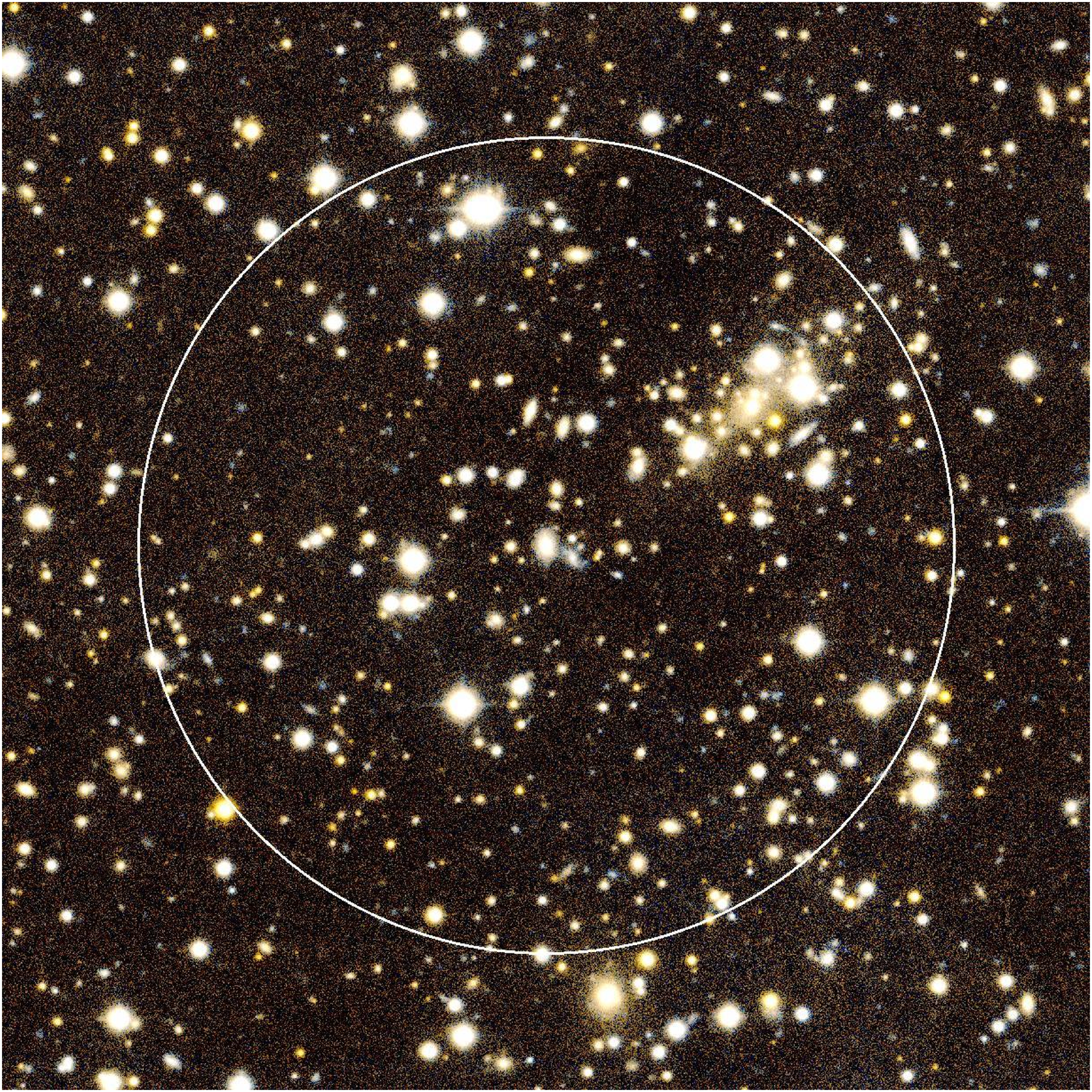}}
\subfigure[\texttt{PSZ2 G042.32+17.48}]{\label{fig:plck-769}\includegraphics[width=0.4\textwidth]{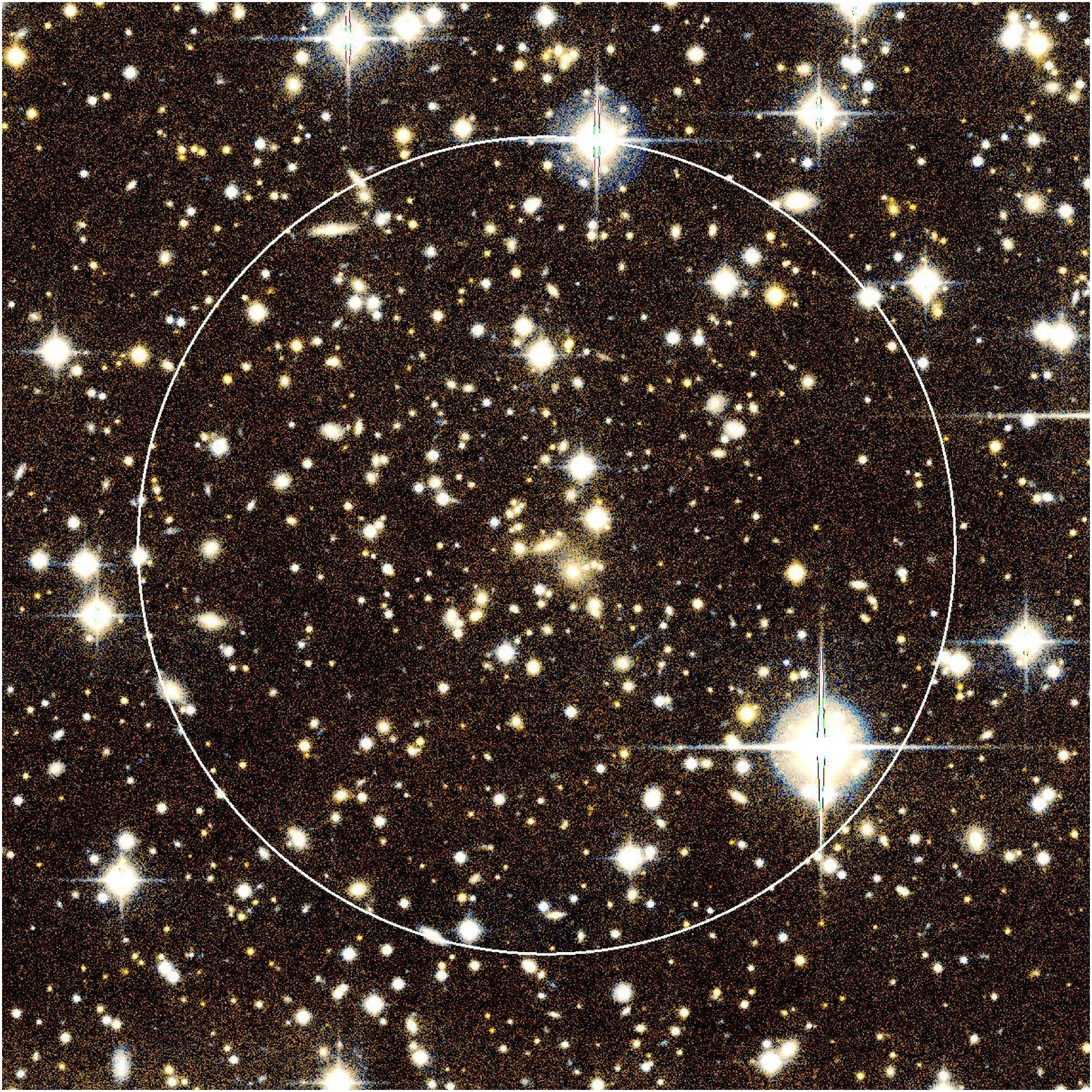}}
\subfigure[\texttt{PSZ2 G048.21-65.00}]{\label{fig:plck-1274}\includegraphics[width=0.4\textwidth]{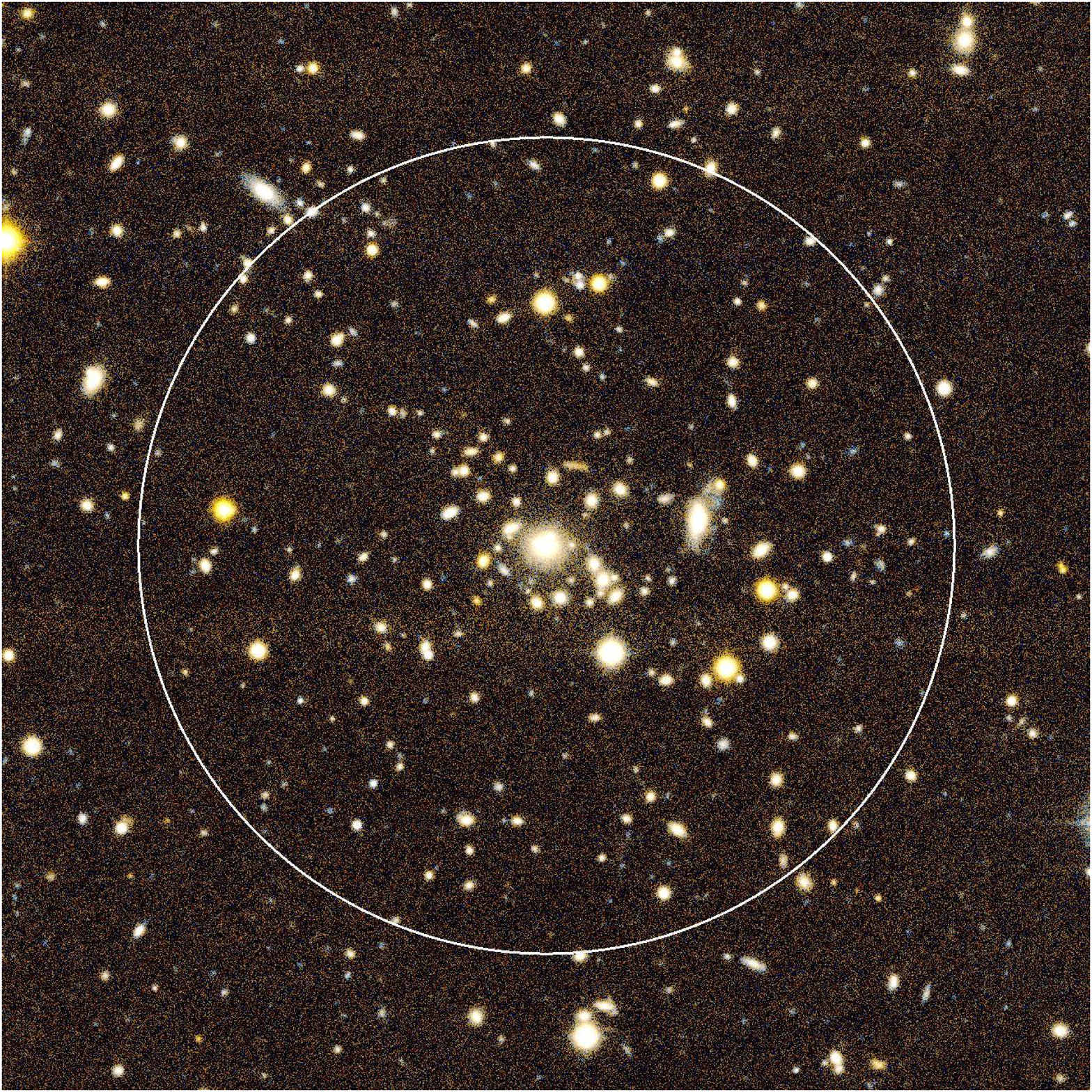}}
\subfigure[\texttt{PSZ2 G071.82-56.55}]{\label{fig:psz2_0356}\includegraphics[width=0.4\textwidth]{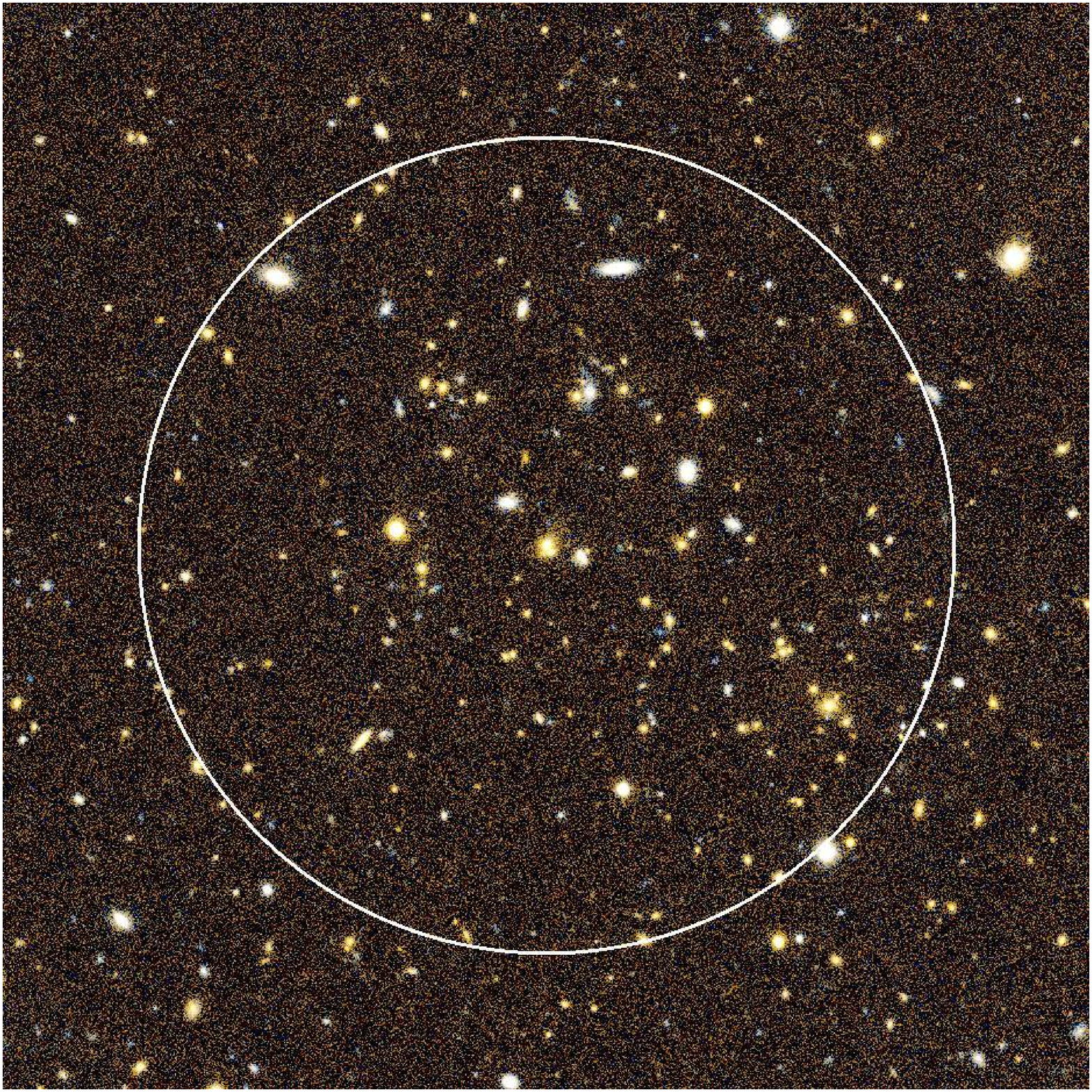}}
\caption{Circles are centred on the position that maximizes the richness, and have a radius of 0.5 Mpc at the estimated redshift. }
\end{figure*}

\begin{figure*}
\centering
\subfigure[\texttt{PSZ2 G076.18-47.30}]{\label{fig:psz2_0377}\includegraphics[width=0.4\textwidth]{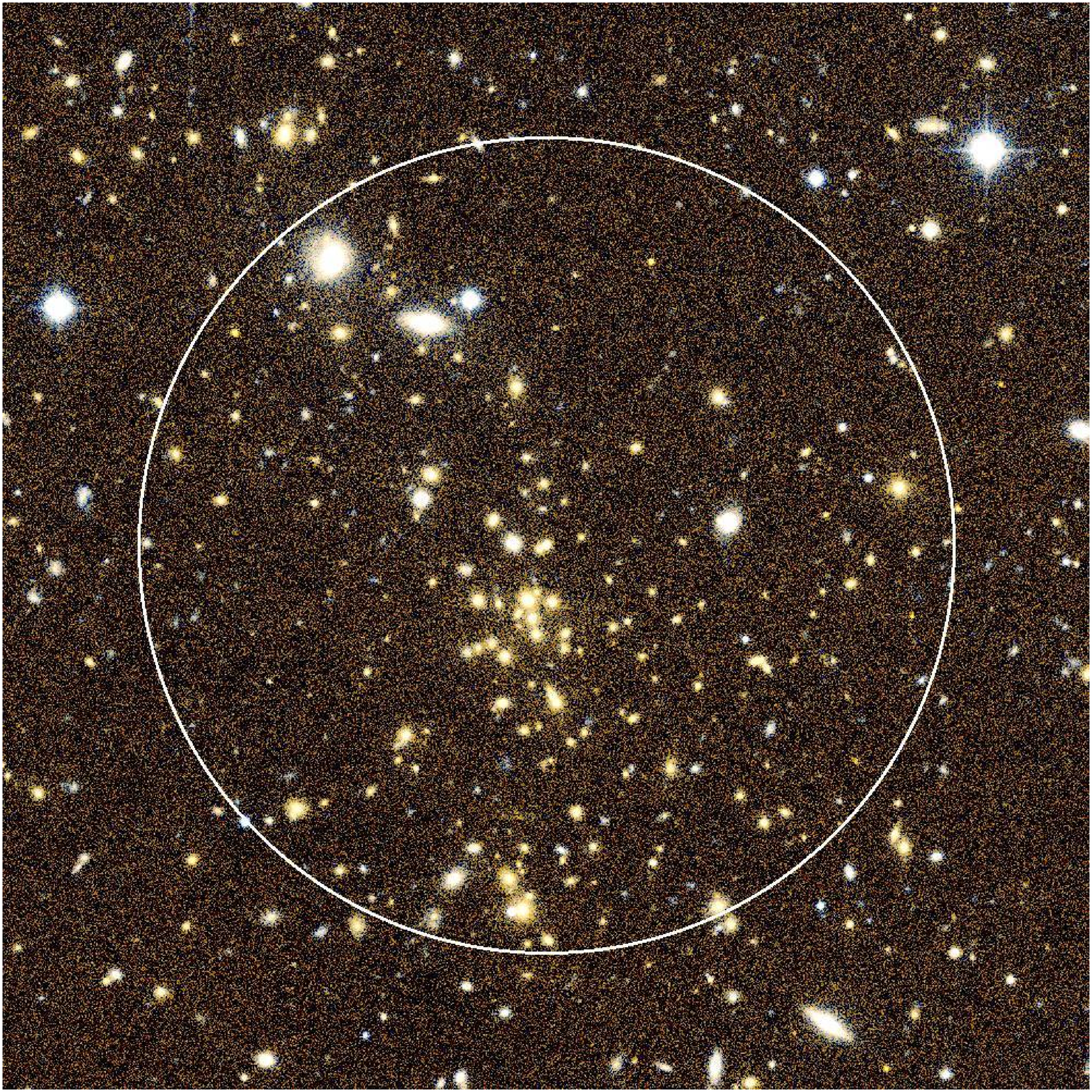}}
\subfigure[\texttt{PLCK G079.95+46.96}]{\label{fig:mmf3-2916}\includegraphics[width=0.4\textwidth]{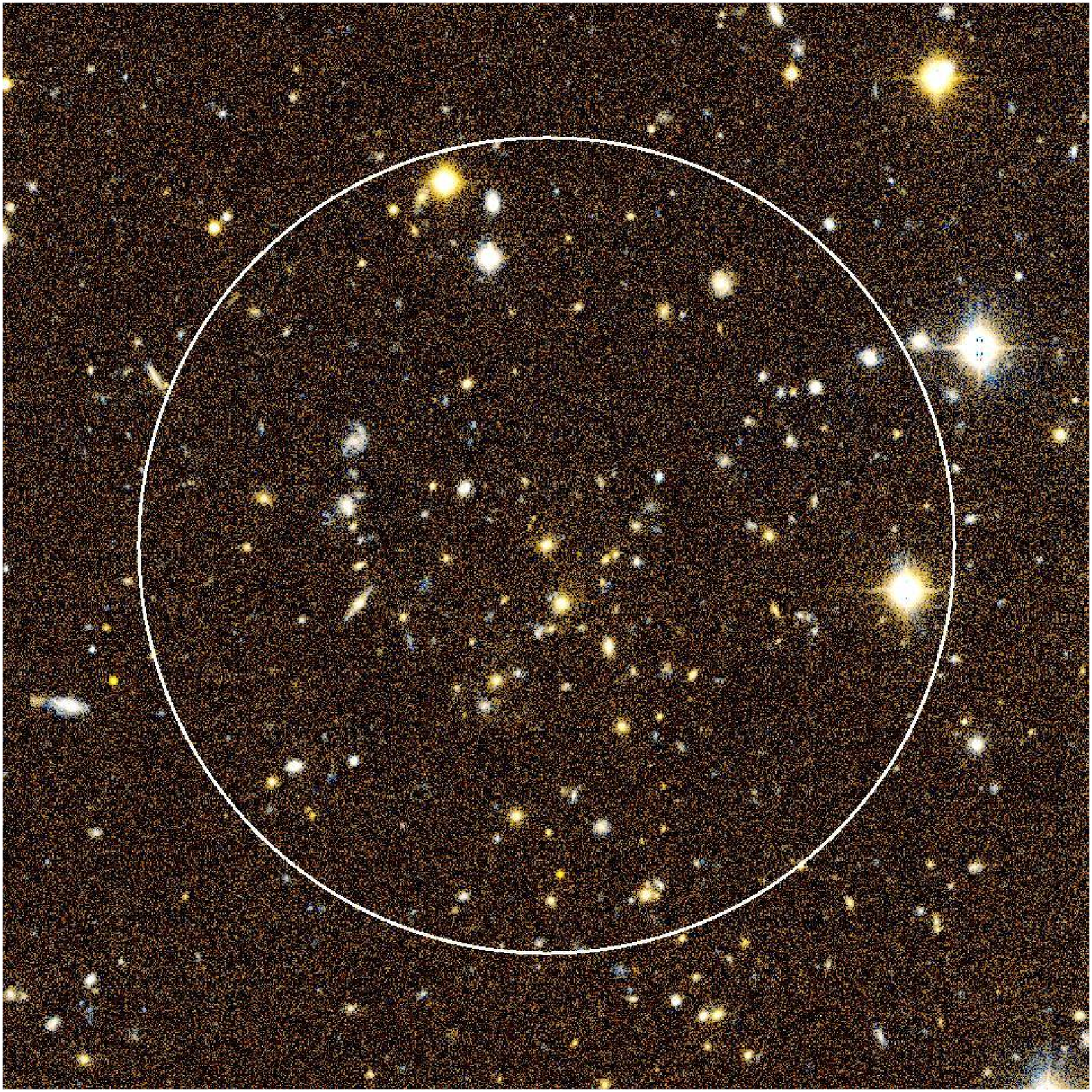}}
\subfigure[\texttt{PLCK G087.58-41.63}: Redshift $z\sim1$ \textit{Planck} cluster. Our choice of filters and the data depth are insufficient to measure a precise redshift and richness for this system.]{\label{fig:psz2_0464}\includegraphics[width=0.4\textwidth]{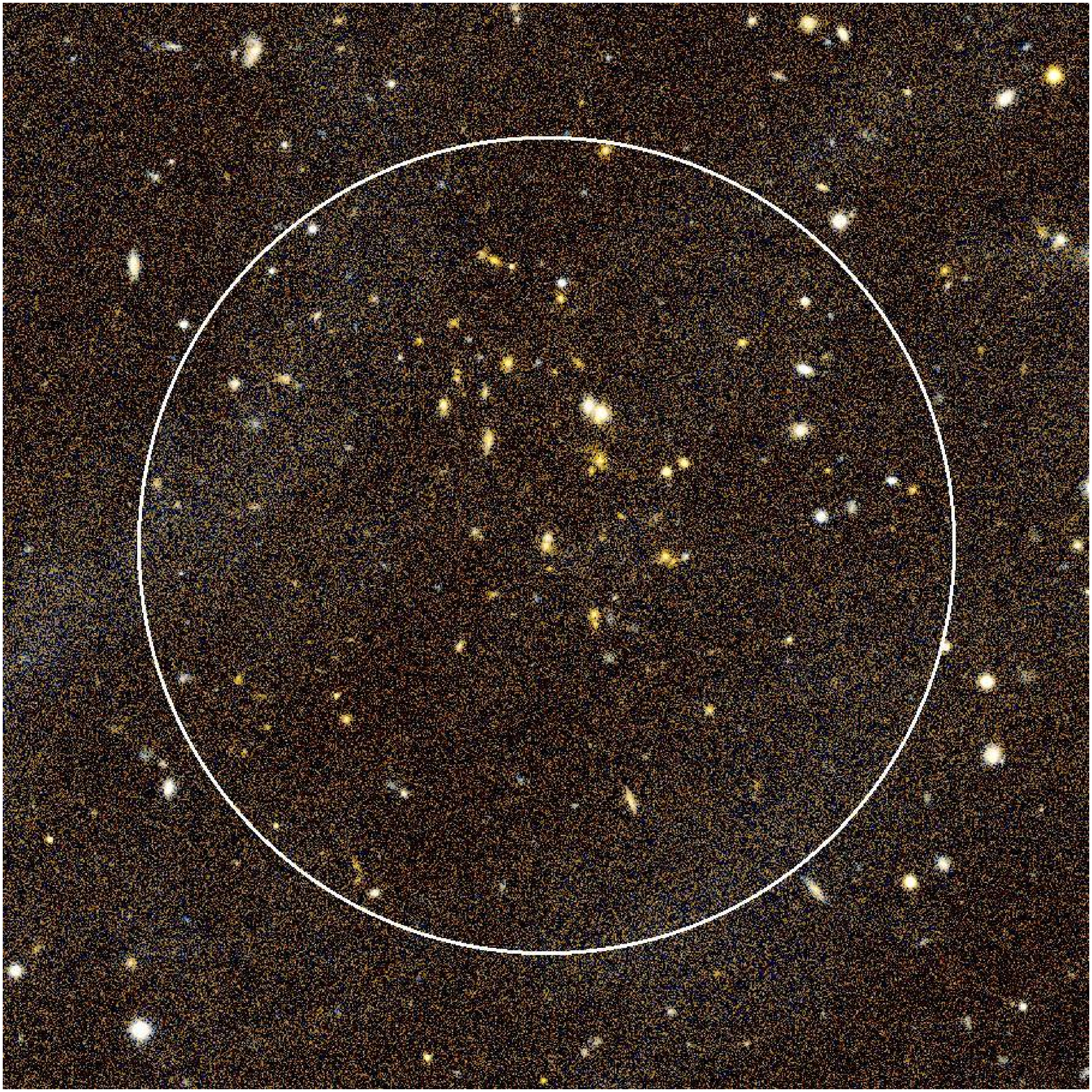}}
\subfigure[\texttt{PSZ2 G106.15+25.75}]{\label{fig:plck-1292}\includegraphics[width=0.4\textwidth]{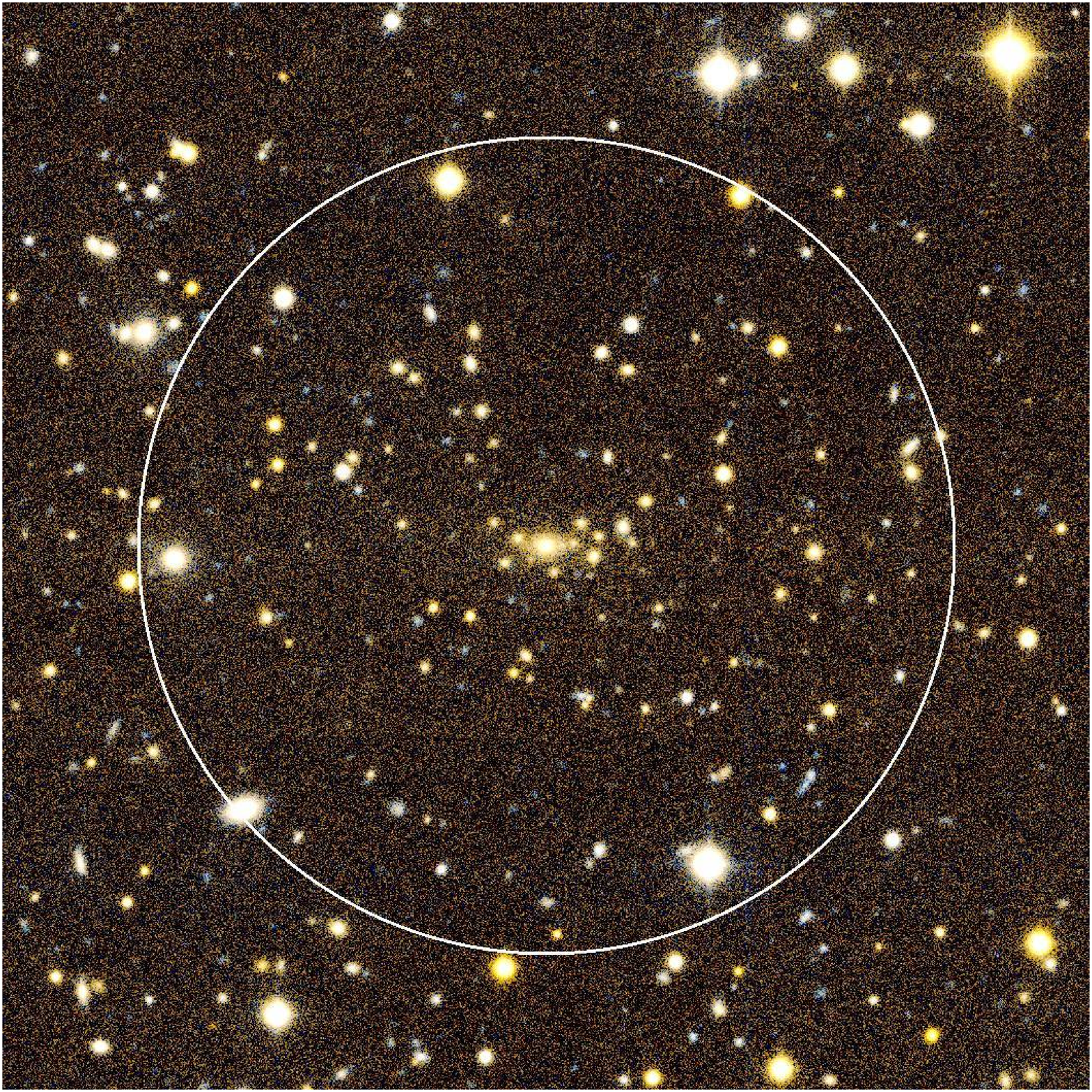}}
\subfigure[\texttt{PSZ2 G119.30-64.68}]{\label{fig:psz2_0693}\includegraphics[width=0.4\textwidth]{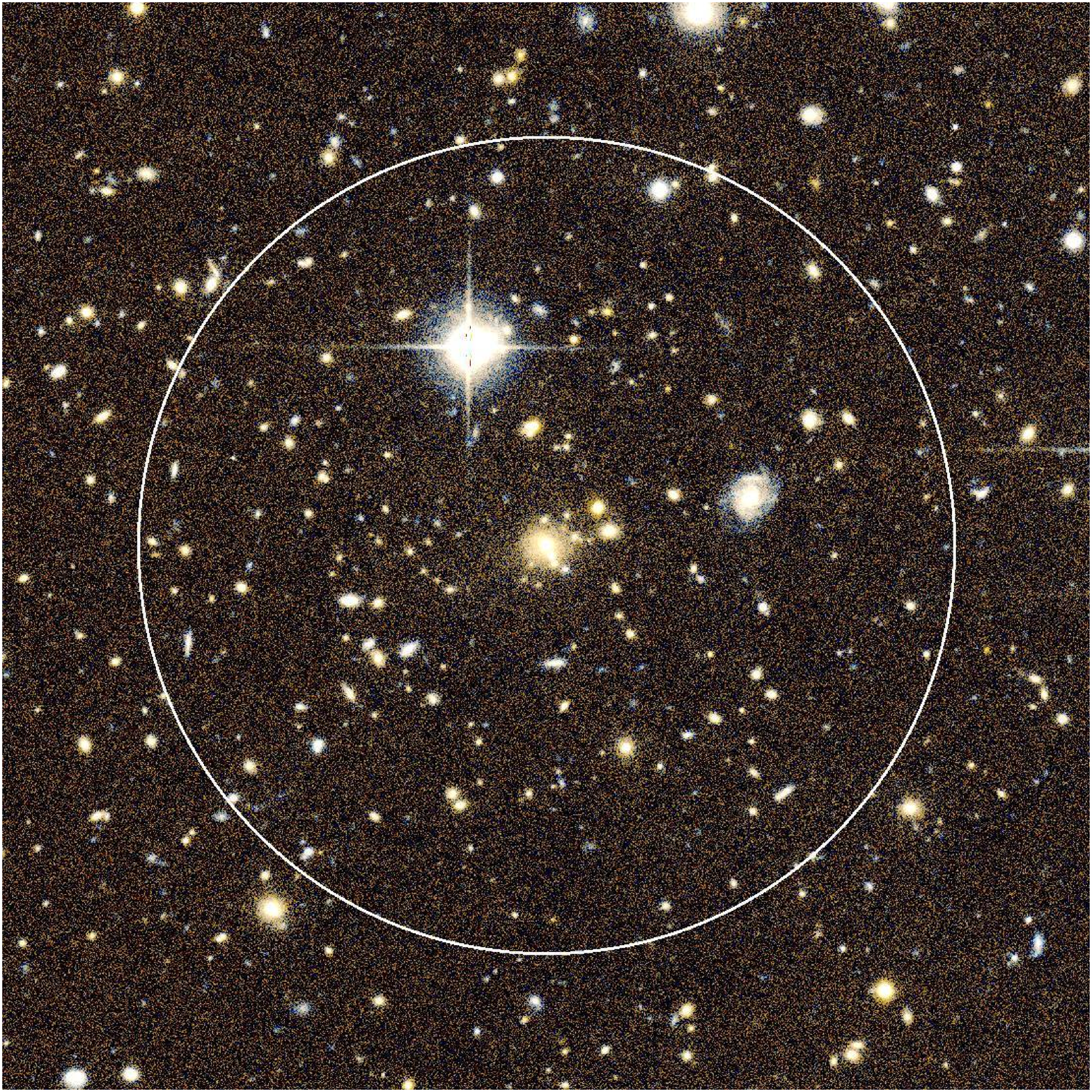}}
\subfigure[\texttt{PSZ2 G141.77+14.19}]{\label{fig:plck-994}\includegraphics[width=0.4\textwidth]{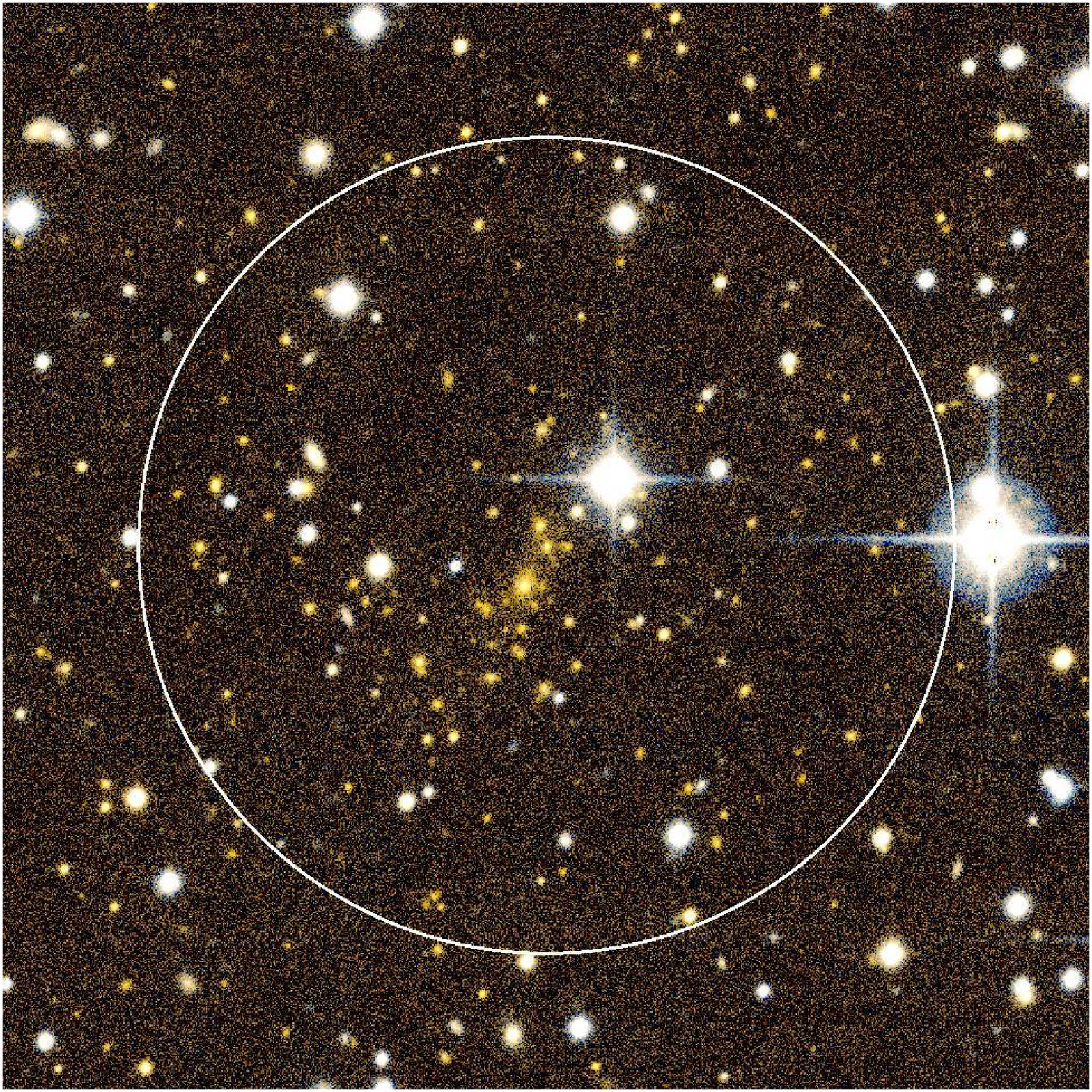}}
\caption{Circles are centred on the position that maximizes the richness, and have a radius of 0.5 Mpc at the estimated redshift. }
\end{figure*}

\begin{figure*}
\centering
\subfigure[\texttt{PLCK G191.75-21.78}]{\label{fig:psz2_1086}\includegraphics[width=0.4\textwidth]{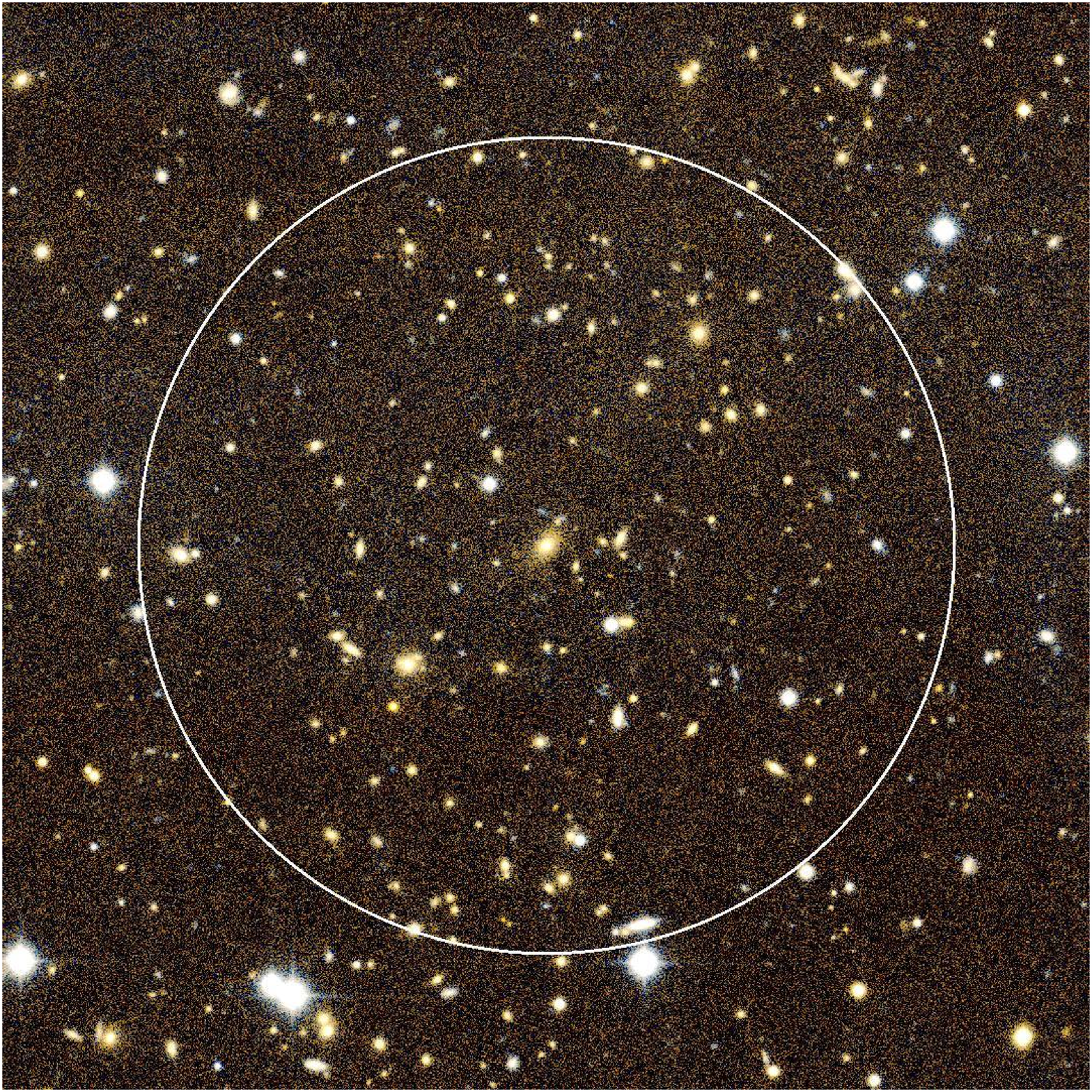}}
\subfigure[\texttt{PSZ2 G198.80-57.57}]{\label{fig:psz2_1112}\includegraphics[width=0.4\textwidth]{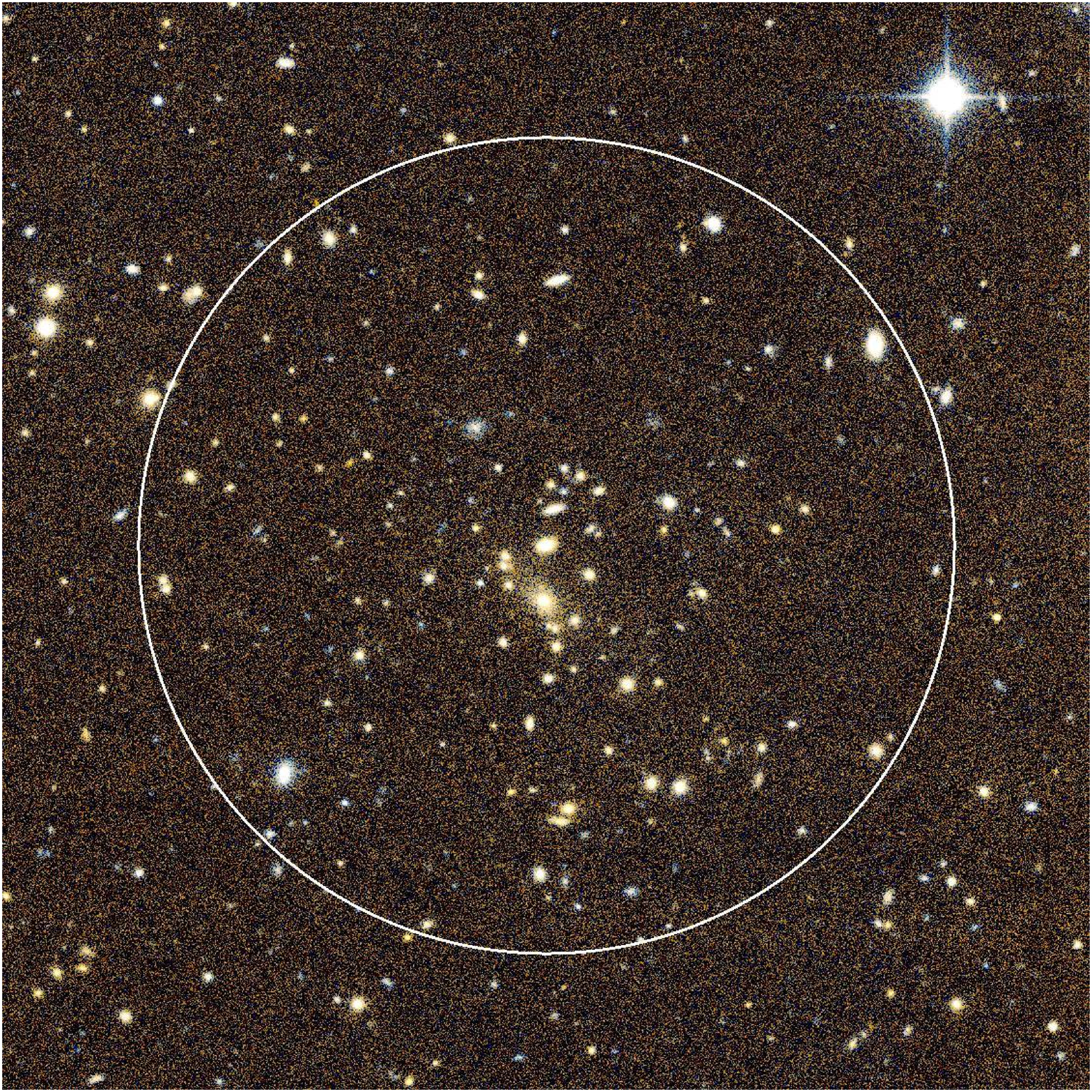}}
\subfigure[\texttt{PSZ2 G208.57-44.31}]{\label{fig:psz2_1158}\includegraphics[width=0.4\textwidth]{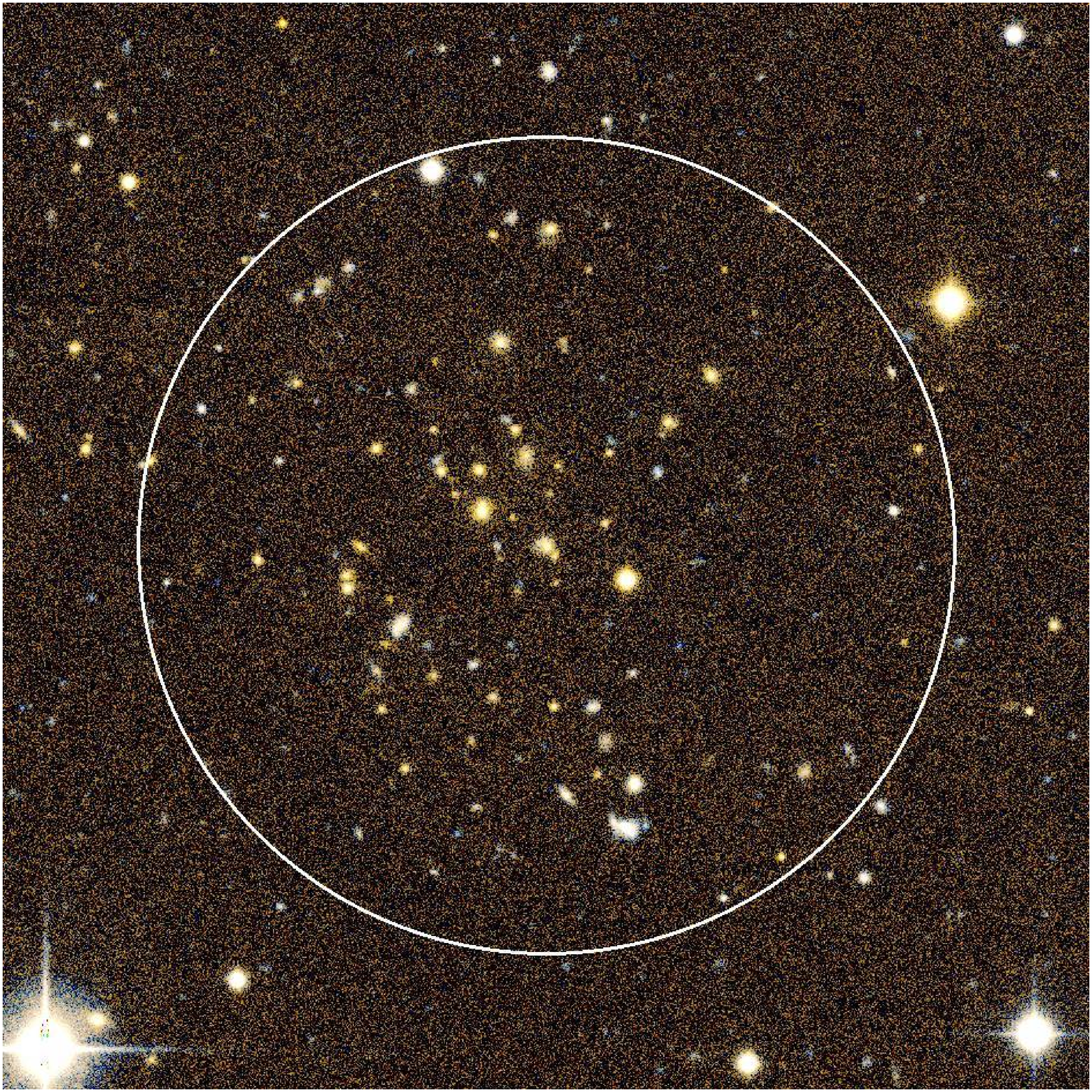}}
\subfigure[\texttt{PLCK G227.99+38.11}]{\label{fig:mmf3-2622}\includegraphics[width=0.4\textwidth]{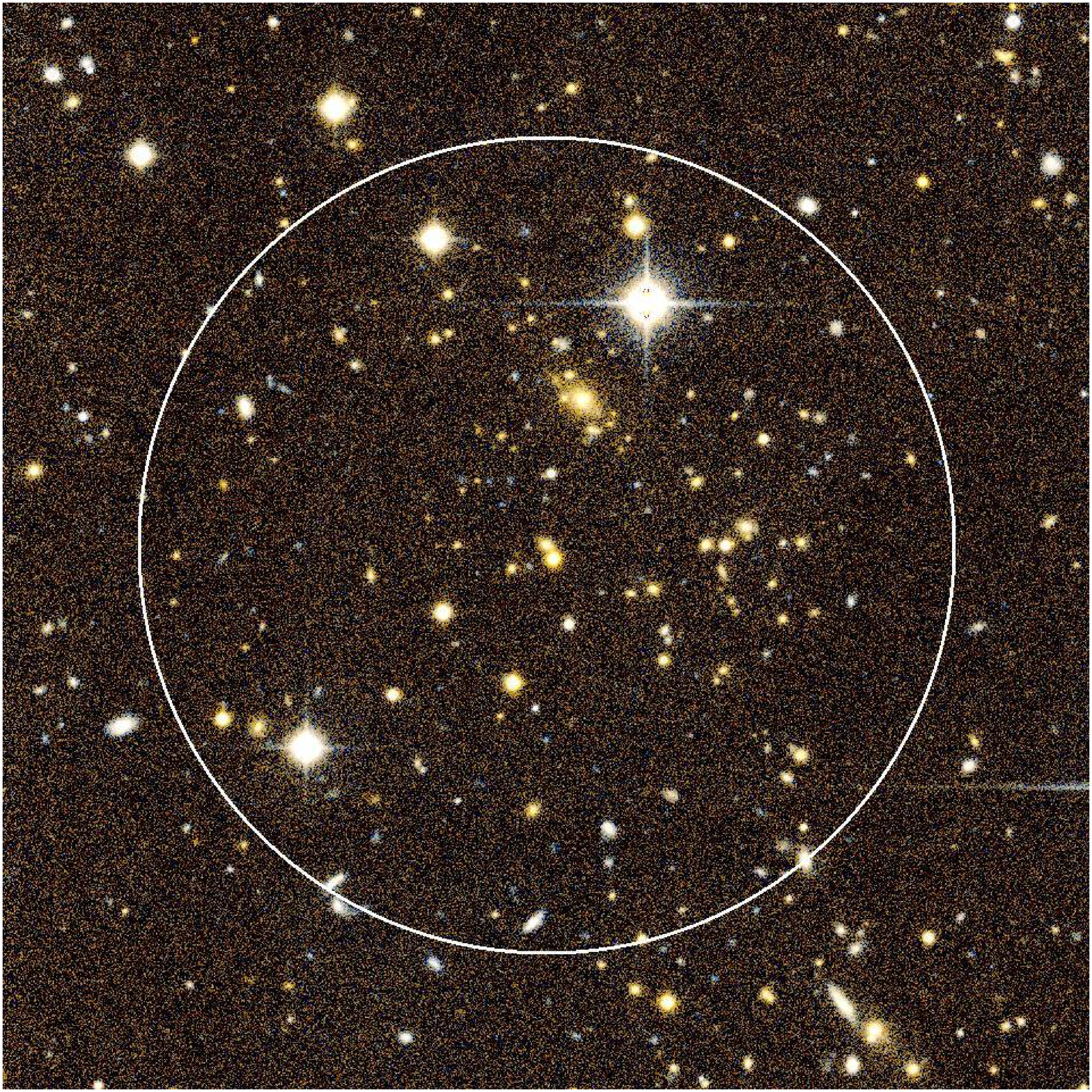}}
\subfigure[\texttt{PSZ1 G038.25-58.36}: Significant overdensity, but not a confirmed \textit{Planck} cluster (cf. Sect.~\ref{sec:candidates}).]{\label{fig:plck-1413}\includegraphics[width=0.4\textwidth]{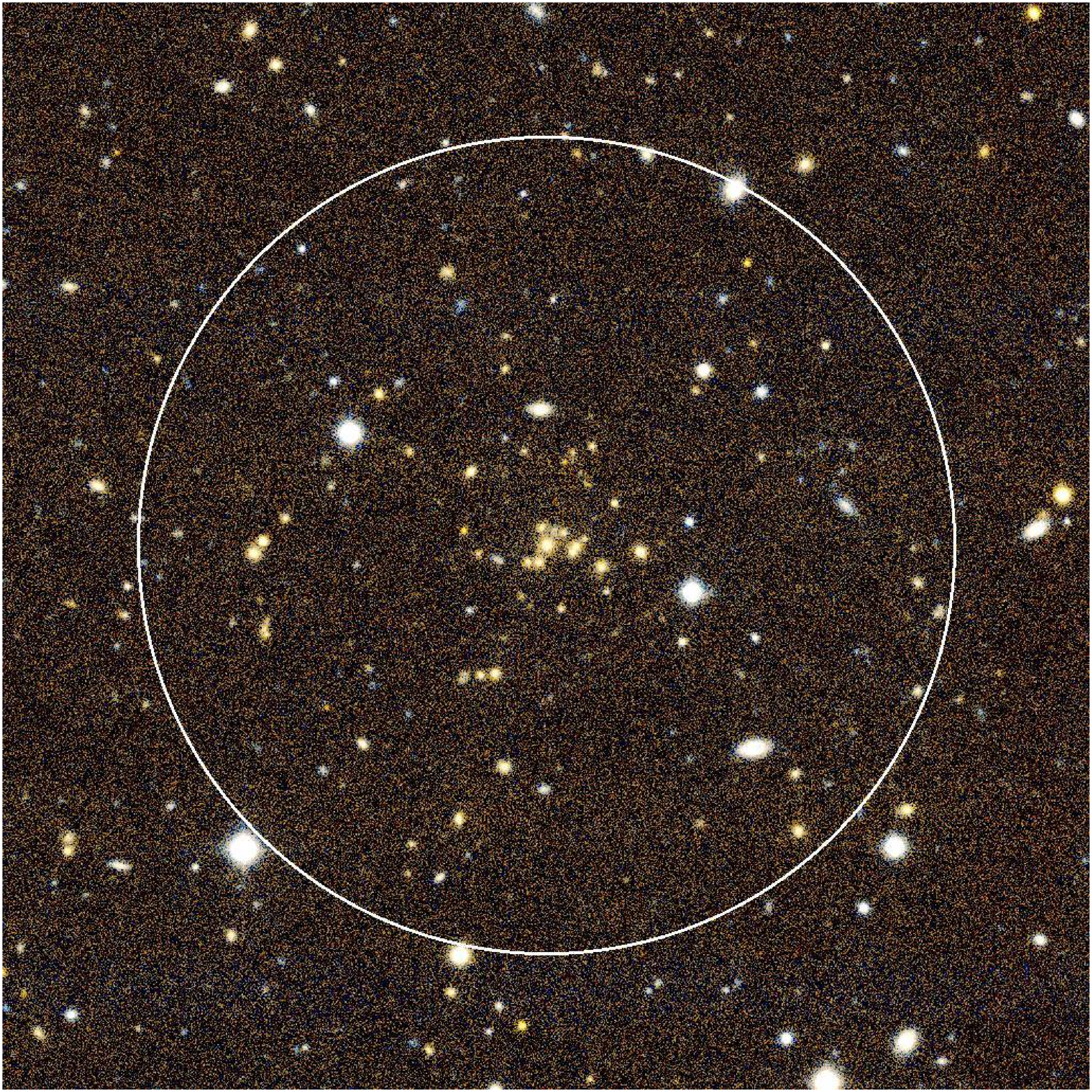}}
\caption{Circles are centred on the position that maximizes the richness, and have a radius of 0.5 Mpc at the estimated redshift. }
\end{figure*}

\end{appendix}
\end{document}